\documentclass[twocolumn,showkeys,showpacs,pra]{revtex4}
\usepackage{amsfonts, amsmath, latexsym}
\usepackage[dvips]{graphicx}

\newtheorem{1}{Theorem}
\newtheorem{2}{Lemma}
\newtheorem{3}[1]{Theorem}
\newtheorem{4}[1]{Theorem}
\newtheorem{5}[1]{Theorem}
\newtheorem{6}[1]{Theorem}

\begin{document}

\title{The Local Variational Principle}

\author{Cristian Predescu}
\email{Cristian_Predescu@brown.edu}
 
\affiliation{
Department of Chemistry, Brown University, Providence, Rhode Island 02912
}
\date{November 15, 2001}

\begin{abstract}
A generalization of the Gibbs-Bogoliubov-Feynman inequality for spinless particles is proven and then illustrated for the simple model of a symmetric double-well quartic potential. The method gives a pointwise lower bound for the finite-temperature density matrix and it can be systematically improved by the Trotter composition rule. It is also shown to produce groundstate energies better than the ones given by the Rayleigh-Ritz principle as applied to the groundstate eigenfunctions of the reference potentials. Based on this observation, it is argued that the Local Variational Principle performs better than the equivalent methods based on the centroid path idea and on the Gibbs-Bogoliubov-Feynman variational principle, especially in the range of low temperatures.  

\end{abstract}


\maketitle

\section{Introduction} 

\newcommand{\ud}{\mathrm{d}}
	The Gibbs-Bogoliubov-Feynman inequality (GBF) is a restatement of the second law of thermodynamics. However, the motivation of the present work is the equally important fact that the inequality provides a variational approximation to the Helmholtz free energy. Historically, Gibbs first stated the inequality for classical systems, then Bogoliubov and Feynman generalized it to quantum systems in the operator and the path-integral formalism of quantum mechanics, respectively.  Perhaps at the expense of losing the original physical significance, the local variational principle I develop in this work is intended to be a mathematical basis for the design of more efficient computational methods dealing with statistical quantum systems, with special concern for their low temperature behavior. To fully justify the need for a local principle, we first have to give a short review of the Feynman and Gibbs-Bogoliubov inequalities and at this point, we shall also introduce some notations of use throughout the paper. 
	
The path integral formulation of the statistical mechanics began with the Feynman's realization at an ``intuitive'' level that the density matrix of a monodimensional quantum particle is the expectation value of a suitable function of a Brownian Motion~\cite{Fey48}. Feynman was actually working on the real time Schr{\"o}dinger equation, but for the imaginary time analog the theory was made rigorous by Ka\c{c}~\cite{Kac50}, the product being the well known Feynman-Ka\c{c} representation formula (Theorem 6.6 of Ref.~\onlinecite{Sim79})
\begin{equation}
\label{eq:1}
\frac{\rho(x,x';\beta)}{\rho_{free}(x,x';\beta)}=\mathbb{E}\exp\left\{-\beta\int_{0}^{1}\! \!  V\Big[x(t)+\sqrt{\frac{\hbar^2\beta}{m}} B_t^0 \Big]\ud t\right\}
\end{equation}
where 
\[x(t)=x+(x'-x)t\]
and
\[\rho_{free}(x,x';\beta)=\sqrt {\frac{m}{2\pi\hbar^2\beta}} \exp{\Big[-\frac{m} {2\hbar^2\beta} (x-x')^2\Big]}\]
is the density matrix for a similar free particle.
$B_t^0$ denotes a standard Brownian Bridge (see p.~40-41 of Ref.~\onlinecite{Sim79} and p.~430-431 of Ref.~\onlinecite{Dur96})
and the expected value in~(\ref{eq:1}) is taken with respect to its underlying probability measure.

For the sake of simplicity, we shall be concerned mainly  with the monodimensional case, but the reader should observe that the theory is in no way restricted to this case. This is so because the Feynman-Ka\c{c} formula has a straightforward multidimensional generalization: one simply utilizes an independent Brownian Bridge for each physical degree of freedom.    However, we explicitly address various multidimensional problems, whenever they significantly differ from their monodimensional version. As stated, the main theorems obtained in this paper remain true for the multidimensional systems. 

The Fourier Path Integral (FPI) implementation of~(\ref{eq:1}), which we exclusively use in this paper, is due to Doll and Freeman~\cite{Dol84} and is based on the exact representation of the Brownian Bridge as a Random Fourier series with the coefficients being independent identically distributed (i.i.d.) Gaussian variables. To rephrase their result in the spirit of the Feynman-Ka\c{c} representation formula, if $\Omega$ is the space of infinite sequences $\bar{a}\equiv(a_1,a_2,\ldots)$ and 
\begin{equation}
\label{eq:2}
P[\bar{a}]=\prod_{k=1}^{\infty}\mu(a_k)
\end{equation}
 is the (unique) probability measure on $\Omega$ such that the coordinate maps $\bar{a}\rightarrow a_k$ are i.i.d. variables with distribution probability
\begin{equation}
\label{eq:3}
\mu(a_k\in A)= \frac{1}{\sqrt{2\pi}}\int_A e^{-z^2/2}\,\ud z
\end{equation}
then,
\begin{equation}
\label{eq:4}
B_t^0(\bar{a})\equiv \sqrt{\frac{2}{\pi^2}}\sum_{k=1}^{\infty}a_k\frac{\sin(k\pi t)}{k},\; 0\leq t\leq1
\end{equation}
is equal in distribution to a standard Brownian Bridge. Let us introduce the path-averaged potential functional
\begin{equation}
\label{eq:5}
U(x,x',\bar{a};\beta)=\int_{0}^{1}V\big[x(t)+\sum_{k=1}^{\infty}a_k \sigma_k \sin(k \pi t)\big]\ud t,
\end{equation}
where
\[\sigma_k^2=\frac{2\beta\hbar^2}{\pi^2 m}\frac{1}{k^2},\]
and make the convention that whenever $x=x'$, the prime $x$ is dropped so that $U(x,\bar{a};\beta)\equiv U(x,x,\bar{a};\beta)$.
With this notation, the FPI version of the Feynman-Ka\c{c} representation formula~(\ref{eq:1}) is
\begin{eqnarray}
\label{eq:6}
  \rho(x,x';\beta)&=&\rho_{free}(x,x';\beta)  \nonumber \\ && \times  \int_{\Omega} \ud P[\bar{a}]\exp\left[-\beta U(x,x',\bar{a};\beta)\right].
\end{eqnarray}

In his treatment of the Fr{\"o}hlich polaron problem~\cite{Fey55}, Feynman constructed an upper bound to the free energy of a quantum system by means of the inequality (see formulae 3.52 and 3.53 in Ref.~\onlinecite{Fey98})
\begin{equation}
\label{eq:7}
F\leq F'_{\bar{b}}+ \langle U-U'_{\bar{b}} \rangle_{U'_{\bar{b}}}
\end{equation}
where, in general, $\langle O \rangle_{S'_{\bar{b}}}$ stands for the average
\begin{equation}
\label{eq:8}
\langle O \rangle_{U'_{\bar{b}}}=\frac{\int_{\mathbb{R}}\ud x \int_{\Omega} \ud P[\bar{a}]e^{-\beta U'_{\bar{b}}(x,\bar{a};\beta)}O(x,\bar{a};\beta)} {\int_{\mathbb{R}} \ud x \int_{\Omega} \ud P[\bar{a}]e^{-\beta U'_{\bar{b}}(x,\bar{a};\beta)}}.
\end{equation}
The functional $U'_{\bar{b}}(x,x',\bar{a};\beta)$ was taken to be of the form~(\ref{eq:5}) for some trial potential $V'_{\bar{b}}(x)$ depending upon a set of parameters $\bar{b}\equiv(b_1,b_2,\ldots)$, but this is not a requirement and essentially any function satisfying some mild integrability conditions can be utilized in the Feynman inequality~(\ref{eq:7}).

As argued by Feynman (see Chapter 11 in Ref.~\onlinecite{Fey65}), the zero temperature limit of the inequality~(\ref{eq:7}) is 
\begin{equation}
\label{eq:9}
\epsilon_{0}\leq \frac{\langle \phi_{\bar{b}}^{0}|\hat{H}| \phi_{\bar{b}}^{0}\rangle}{\langle \phi_{\bar{b}}^{0}|\phi_{\bar{b}}^{0}\rangle},
\end{equation}
where $\phi_{\bar{b}}^{0}$ is the groundstate eigenfunction of the trial Hamiltonian
\begin{equation}
\label{eq:10}
\hat{H}'_{\bar{b}}=-\frac{\hbar^{2}}{2m}\frac{\partial^{2}}{\partial x^{2}}+V'_{\bar{b}}(x).
\end{equation}
Eq.~\ref{eq:9} is in agreement with the Rayleigh-Ritz principle for groundstate eigenfunctions and, as an approximation, can be arbitrarily sharpened by use of more accurate trial potentials. By the inherent continuity of such problems, these good variational estimates of the groundstate energy imply good estimates of the Helmholtz free energy for the entire low temperature regime, fact hard to achieve by other means. This helps explain the successful application of the Feynman variational principle in a variety of theories dealing with the evaluation of the thermodynamic properties of quantum systems~\cite{Gia85, Fey86, Lob92, Cuc95, Hwa97, Kle01}. 

The Gibbs-Bogoliubov inequality~\cite{Kva56, Rue69} provides the following bound to the free energy
\begin{equation}
\label{eq:11}
F\leq F'_{\bar{b}}+ \frac{Tr[(\hat{H}-\hat{H}'_{\bar{b}})e^{-\beta \hat{H}'_{\bar{b}}}]}{Tr (e^{-\beta \hat{H}'_{\bar{b}}})},
\end{equation}
which for spinless particles is proven to be equal to the one given by the Feynman inequality~\cite{Fey98}, whenever the functional $U'_{\bar{b}}(x,x',\bar{a};\beta)$ can be cast in the form of the equation~(\ref{eq:5}) for a given trial potential $V'_{\bar{b}}(x)$. In this situation one talks about the \emph{Gibbs-Bogoliubov-Feynman inequality (GBF)} and of the corresponding variational principle consisting of the minimization of the right-hand expressions in the formulae~(\ref{eq:7}) and~(\ref{eq:11}) on the set of parameters $\bar{b}$. The reader should not conclude that the Gibbs-Bogoliubov inequality is automatically weaker than its path-integral counterpart. For instance, in the case of a fermionic system, the Gibbs-Bogoliubov inequality is still true if the trace is restricted to the Hilbert space of antisymmetric functions, with the slight requirement that the trial Hamiltonian $\hat{H}'_{\bar{b}}$  be totally symmetrical under the permutation of identical particles. However, there is no known path-integral equivalent to the resulting inequality, the difficulty being related to the so-called fermionic sign problem~\cite{Cep91}. It is for this reason that we shall restrict the development of our local variational principle to spinless systems. 

	The GBF usefulness depends upon our ability to analytically compute the integrals on the right-hand side of the equation~(\ref{eq:7}), at least the ones with respect to the Fourier coefficients. This effectively restricts the choice of trial potentials to a handful (in most cases a quadratic potential) and it is in poor match with the fact that the Feynman estimate is global, involving an integration over the physical coordinates. We thus arrive at the first motivation for our work: a local fitting (pointwise in the configuration space), as opposed to a global one, would make more use of a simple reference potential. Then, a pointwise approximation of the density matrix can always be improved by other means than the use of more complicated trial potentials, the default choice being the Trotter composition rule~\cite{Tro59}. Finally, we will show that the local variational principle developed in this work provides new information  about the density matrix, information unattainable from GBF. Our perspective on the computation of the density matrix is thus changed: instead of seeking better reference potentials, we try to find the variational principle which makes the best use of a given reference potential.  
 
\section{The   Local Variational Principle as a lower bound for the finite temperature density matrix}

The purpose of this section is to define the Local Variational Principle (LVP) and further justify its importance. In addition, we shall consider the particular case of LVP when the reference potential is the quadratic one and compare this case with the centroid based approximations~\cite{ Fey86, Gia86, Vot89, Cao90, Vot90, Cao94}, particularly with EFLT~\cite{Cao90}, which is the GBF analog.

The Gibbs-Bogoliubov-Feynman inequality is a consequence of Jensen's inequality  and I remind the reader the latter's statement (see p.~14 in Ref.~\onlinecite{Dur96}):
\begin{1}[Jensen's inequality]
If \mbox{$(\Omega,P)$} is a probability space, if $g:\Omega \rightarrow \mbox{(a,b)}$ is integrable and if $F$ is convex on $\mbox{(a,b)}$ with $-\infty \leq a < b \leq \infty$, then \[\int F\circ g\,\ud P \geq F\left( \int g \,\ud P \right).  \] 
\end{1}

By default, whenever we apply Jensen's inequality in this work, it is understood that the convex function is the exponential $F(x)=\exp(-x)$.

To begin with the definition of the Local Variational Principle, let us perform a change of measure in~(\ref{eq:6}) of the form
\begin{eqnarray}
\label{eq:12}
 \rho(x,x';\beta)=\rho_{free}(x,x';\beta)\int_{\Omega}\, \ud P[\bar{a}] e^{-\beta U'_{\bar{b}}(x,x',\bar{a};\beta)}  \nonumber \\ \times \exp \{-\beta  \left[U(x,x',\bar{a};\beta)-U'_{\bar{b}}(x,x',\bar{a};\beta)\right]\}, \quad 
\end{eqnarray}
where $U'_{\bar{b}}(x,x',\bar{a};\beta)$ is any measurable function depending upon a set of parameters $\bar{b}=(b_1,b_2,\ldots)$ such that 
\begin{equation}
\label{eq:13}
\rho'_{\bar{b}}(x,x';\beta)=\rho_{free}(x,x';\beta)\int_{\Omega}\ud P[\bar{a}] e^{-\beta U'_{\bar{b}}(x,x',\bar{a}; \beta)}
\end{equation}
has an integrable diagonal. Defining a new probability measure by the relation
\begin{equation}
\label{eq:14}
\ud P'_{(x,x',\bar{b};\beta)}[\bar{a}]=\frac{\rho_{free}(x,x';\beta)}{\rho'_{\bar{b}}(x,x';\beta)} e^{-\beta U'_{\bar{b}}(x,x',\bar{a};\beta)}\ud P[\bar{a}],
\end{equation}
we may rewrite~(\ref{eq:12}) as
\begin{eqnarray}
\label{eq:15}
 \rho(x,x';\beta)=\rho'_{\bar{b}}(x,x';\beta) \int_{\Omega}\, \ud P'_{(x,x',\bar{b};\beta)}[\bar{a}] \nonumber  \\ \times \exp\left\{{-\beta \left[U(x,x',\bar{a};\beta)-U'_{\bar{b}}(x,x',\bar{a};\beta)\right]}\right\}.
\end{eqnarray}
Now, use of the Jensen's inequality produces the \emph{local variational inequality}
\begin{equation}
\label{eq:16}
\rho(x,x';\beta)\geq \rho_{\bar{b}}^{a}(x,x';\beta),
\end{equation}
where
\begin{eqnarray}
\label{eq:17}
\rho_{\bar{b}}^{a}(x,x';\beta)&=& \rho'_{\bar{b}}(x,x';\beta) \exp \Big\{-\beta\int_{\Omega}\, \ud P'_{(x,x',\bar{b};\beta)}[\bar{a}] \nonumber \\&& \times {\left[U(x,x',\bar{a};\beta)-U'_{\bar{b}}(x,x',\bar{a};\beta)\right]}\Big\}.
\end{eqnarray}
This inequality is the cornerstone of the variational methods, providing a bound from below to the density matrix. Specialized versions of the inequality were considered before whether as a starting point for the definition of the Partial Averaging method \cite{Dol85} or in the context of the Feynman-Kleinert variational-perturbational theory \cite{Bac99, Kle99}. We should remark here that the nonnegativity of the density matrix, which stems from the reality of the path-averaged potential functional, played an important role. Therefore, the inequality is not true for general complex $\beta$. The local variational inequality~(\ref{eq:16}) implies the Feynman inequality~(\ref{eq:7}). The latter can be deduced by setting $x=x'$ in~(\ref{eq:16}), integrating over $x$, working along the same lines as in~(\ref{eq:14}-\ref{eq:15}), and finally using again Jensen's inequality to obtain
\begin{eqnarray}
\label{eq:18}
e^{-\beta F}&\geq& e^{-\beta F'_{\bar{b}}}\int_{\mathbb{R}} \ud x \frac{\rho'_{\bar{b}}(x;\beta)} {e^{-\beta F'_{\bar{b}}}} \exp \Big\{-\beta\int_{\Omega}\, \ud P'_{(x,\bar{b};\beta)}[\bar{a}] 
\nonumber \\&&\times \left[U(x,\bar{a};\beta)-U'_{\bar{b}}(x,\bar{a};\beta)\right]\Big\}
\nonumber \\ &\geq& 
e^{-\beta F'_{\bar{b}}}\exp \left[-\beta \langle U-U'_{\bar{b}} \rangle_{U'_{\bar{b}}} \right],
\end{eqnarray}
which produces~(\ref{eq:7}) upon taking the logarithm.

For the rest of the paper, we shall only be concerned with the case when the functional $U'_{\bar{b}}(x,x',\bar{a};\beta)$ is the path average of some reference potential depending upon the set of parameters $\bar{b}$
\begin{equation}
\label{eq:19}
U'_{\bar{b}}(x,x',\bar{a};\beta)=\int_{0}^{1}V'_{\bar{b}} \big[x(t)+ \sum_{k=1}^{\infty}a_k \sigma_k \sin(k \pi t) \big]\ud t.
\end{equation}
The reference potential $V'_{\bar{b}}$ is assumed to be a \emph{bounding} potential, with a discrete spectrum and a unique and strictly positive groundstate eigenfunction. We shall denote its eigenfunctions by~$\phi^k_{\bar{b}}(z)$ and the corresponding eigenvalues by~$\epsilon^k_{\bar{b}}$.
Taking the supremum in the equation~(\ref{eq:16}) over the set of parameters~$\bar{b}$ produces the sharper \emph{  Local Variational Principle} (LVP)
\begin{equation}
\label{eq:20}
\rho(x,x';\beta)  \geq  \rho_{best}(x,x';\beta) \equiv \sup_{\bar{b}}\rho_{\bar{b}}^{a}(x,x';\beta). 
\end{equation}
We say that $\rho_{best}(x,x';\beta)$ is the best approximation of the density matrix in the sense of LVP\@. The extremum of the maximization problem~(\ref{eq:20}) is attained on some parameters $\bar{b} \equiv \bar{B}(x,x',\beta)$ which generally are functions of position and temperature (in case of multiple maxima, choose arbitrarily one of them) and as a direct consequence, $\rho_{best}(x,x';\beta)$ is no longer the density matrix of a trial potential. 

At this point, it is useful to consider the special case of LVP when the reference potential is the harmonic oscillator one.  The method will be termed HO-LVP. I only present the monodimensional version as a clear suggestion of how the inequality~(\ref{eq:20}) can be employed. The multidimensional version as well as the specific numerical implementation and the related problems will make the object of a separate paper. The most general monodimensional quadratic potential has the form:
\begin{equation}
\label{eq:20a}
V'_{\omega}(x)=\frac{1}{2}m_0\omega^2(x-z)^2
\end{equation}
where the translational variable~$z$ and the frequency~$\omega$ are the parameters to be determined by the local variational principle. Straightforward but lengthy calculations by means of the FPI path integral formulation, give the following HO-LVP approximation for the density matrix:
\begin{widetext}
\begin{eqnarray}
\label{eq:20aa}
\frac{\rho^a_{z,\omega}(x,x';\beta)}{\rho_{free}(x,x';\beta)}=h_0(\beta C)  \exp{\Big[-\frac{1}{2}\beta^3B^2h_5(\beta C)-\frac{1}{2}\beta^3A^2h_6(\beta C) +\frac{1}{2}\beta^2C^2h_4(\beta C)\Big]} 
\nonumber \\ \times
 \exp{ \Bigg\{ -\beta\int_{0}^{1}\!\! \ud t \; \overline{V}_{t,\omega} \Big[x(t)-}  
 \sqrt{\frac{2\hbar^2}{\pi^2 m}}\beta^2B h_2(\beta C,t)-\sqrt{\frac{2\hbar^2}{\pi^2 m}}\beta^2A h_3(\beta C,t) \Big] \Bigg\}. \quad
\end{eqnarray} 
\end{widetext}

In the above,
\begin{eqnarray}
\label{eq:20b}
C & = & \frac{\hbar \omega}{\pi} \nonumber \\ 
B & = & \sqrt{\frac{2\hbar^2}{\pi^2 m}}\frac{m\omega^2}{\pi}(x+x'-2z)  \\
A & = & \sqrt{\frac{2\hbar^2}{\pi^2 m}}\frac{m\omega^2}{\pi}(x-x'). \nonumber
\end{eqnarray}
Even though $A$, $B$ and $C$ are in fact functions of $\omega$, $z$, $x$ and $x'$, we do not write their arguments explicitly in order to save typographical space. The h-functions are tabled in Appendix~A. In addition, 
\begin{equation}
\label{eq:20c}
\overline{V}_{t,\omega}(x)=\int_{\mathbb{R}}  V(x+z)\frac{1}{\sqrt{2\pi \Gamma_{\omega}^2(t)}} \exp\left[-\frac{z^2}{2\Gamma_{\omega}^2(t)}\right] \ud z
\end{equation}
is a convolution of the original potential with a Gaussian of width
\begin{equation}
\label{eq:20d}
\Gamma_{\omega}^2(t)=\frac{2\beta \hbar^2}{\pi^2 m}h_1(\beta C,t) 
\end{equation}
and is called a Gaussian transform of the potential.
	
	The inequality~(\ref{eq:20}) simply states that
\[
\rho(x,x';\beta)\geq \rho^a_{z,\omega}(x,x';\beta)
\]
so that, for each pair of points $(x,x')$ the maximum of the right-hand side expression is attained on some optimum values of the parameters $z=Z(x,x';\beta)$ and $\omega=\Omega(x,x';\beta)$. 

I do not discuss here how the HO-LVP technique can be numerically implemented for practical applications. With the sole difference that there are monodimensional integrals against the parameter~$t$ to be computed numerically  (a tractable problem), the HO-LVP is on par as computational difficulty with the centroid based methods~\cite{Vot89, Cao90, Vot90, Cao94} and it is amenable to similar approximations (see Ref.~\onlinecite{Lob92} for an example). They all involve local minimizations and integrations in the configuration space, respectively in the centroid space.  

Rather, we shall emphasize the differences between such methods. Since it is required that the density matrix of the reference potential be analytically known, a common feature of the methods is the fact that only simple references, as for instance the quadratic potential reference, are computationally feasible. Therefore, it is desirable that the approximation which makes the better use of the simple reference potential be employed in actual simulations. As such, HO-LVP has two important advantages over EFLT~\cite{Cao90}: (a) it can be arbitrarily improved by Trotter composition~\cite{Tro59} and (b) it gives an approximation to the true density matrix which provides more information about the system than the variational approximation to the centroid density matrix.  

Because the first property is clear, we shall be mainly concerned in this paper with proving the second assertion. To this point, we notice that the high temperature limit of any density matrix is the classical one. However, as the temperature is lowered, the discrepancy between the classical and the quantum density matrices increases because the thermodynamic spread of the paths entering the Feynman-Ka\c{c} formula also increases. In fact, if the number of variables used to parameterize the paths is kept constant, the thermodynamic energy estimator for non-variational methods as DPI~\cite{Str77, Sch81, Cha81, Thi84} or FPI~\cite{Dol84} extrapolates to the classical energy in the low temperature limit too~\cite{Kau84}, a phenomenon dubbed ``classical  collapse.''  Giachetti and Tognetti \cite{Gia85} as well as Feynman and Kleinert~\cite{Fey86} noticed that this is not true of the variational methods based upon the GBF principle.  Following their line of thought, one may argue that the low temperature limit of the energy estimator for the EFLT centroid method (see the equations~2.34, 2.35, 2.41, and~2.42 of Ref.~\onlinecite{Cao90}) is
\begin{equation*}
\inf_{z,\omega} \frac{\langle \phi_{z,\omega}^{0}|\hat{H}| \phi_{z,\omega}^{0}\rangle}{\langle \phi_{z,\omega}^{0}|\phi_{z,\omega}^{0}\rangle}\geq \epsilon_0,
\end{equation*}
i.e, the expected energy of the best Gaussian wavepacket. [Remember that $\phi_{z,\omega}^{0}(x)$ is the groundstate eigenfunction of the reference potential given by Eq.~\ref{eq:20a}]. 
This is a quite remarkable fact because the harmonic oscillator is known to be a good approximation of the potential surface around the main local minima.

However, one very important aspect of the centroid density matrix $\rho^c(\bar{x}; \beta)$ is that it bears no direct connection to the true density matrix. For instance, given $\rho^c(\bar{x};\beta)$, one cannot compute exactly the ensemble average potential energy
\[
\frac{\int_{\mathbb{R}} \rho(x;\beta) V(x) \ud x}{\int_{\mathbb{R}} \rho(x;\beta) \ud x},
\]
though useful approximations are known~\cite{Vot90}. As far as the total energy is concerned, this can be exactly evaluated with the help of the T-method estimator (see Section~V for definition). But up to some functionals of it as for example the partition function, this is the \emph{only  property} that can be computed exactly once the centroid density matrix is known. Clearly, there is no H-method estimator  for the centroid density matrix. Things are totally different in the case of the Local Variational Principle because this provides an approximation for the true density matrix and so, the expectation values of different operators are readily available. Even more, we shall later show that the zero temperature limit of the H-method estimator is a groundstate energy estimate \emph{always} better than the corresponding centroid one, the latter being matched by the low temperature limit of the LVP T-estimator. I hope this would be enough evidence to convince the reader that LVP provides the better description of the physical system.  

From the above discussion, we infer that the quality of a variational approximation is dictated by its low temperature limit, and in the next section we shall establish what this limit is in the case of the Local Variational Principle. The reader should not forget that LVP provides a variational bound from below to the \emph{finite-temperature} density matrix and that it is intended as an approximation method for this density matrix. Therefore, LVP is in no way limited to the computation of the groundstate eigenfunction, which is however our object of interest for the next section.    

\section{The eigenfunction representation of the LVP}

	In order to establish the low temperature limit of the Local Variational Principle, as well as to show that the local variational inequality~(\ref{eq:20}) can also be interpreted as a generalization of the Gibbs-Bogoliubov inequality, we need to express $\rho_{\bar{b}}^{a}(x,x';\beta)$ in terms of the eigenfunctions and the eigenvalues of the potential $V'_{\bar{b}}(x)$. We again develop the theory in full generality, rather than discussing the special HO-LVP case. 
	
	In anticipation of the final result, let us see that the eigenfunctions and the eigenvalues of the perturbed Hamiltonian  
\begin{equation}
\label{eq:21}
\hat{H}'_{\bar{b},\lambda}= \hat{H}'_{\bar{b}} + \lambda \left[V(x)-V'_{\bar{b}}(x)\right]
\end{equation}
as given by the first order Rayleigh-Schr{\"o}dinger perturbation theory are of the form
\begin{equation}
\label{eq:22}
\phi_{\bar{b}, \lambda}^k (x)\approx \phi_{\bar{b}}^k (x) - \lambda \sum_{i\neq k}^{\infty} c_{ki}\phi_{\bar{b}}^{i}(x) 
\end{equation}
respectively,
\begin{equation}
\label{eq:23}
\epsilon^k_{\bar{b}, \lambda} \approx \epsilon^k_{\bar{b}} + \lambda {\left\langle{\phi_{\bar{b}}^k} |V-V'_{\bar{b}}| {\phi_{\bar{b}}^k}\right\rangle},
\end{equation}
provided that the eigenfunctions $\phi_{\bar{b}}^{i}(x)$ are chosen such that the perturbation $V(x)-V'_{\bar{b}}(x)$ is diagonalized on each degenerate subspace. Letting $D_{k}=\{i \in \mathbb{N}:\epsilon^i_{\bar{b}}= \epsilon^k_{\bar{b}} \}$ and $\nu_{ki}=\langle{\phi_{\bar{b}}^k} |V-V'_{\bar{b}}| {\phi_{\bar{b}}^i}\rangle$, the exact expressions for the coefficients $c_{ki}$ are (see Chapter 5 in Ref.~\onlinecite{Sak94}):
\begin{eqnarray}
\label{eq:24}
c_{ki}&= &\frac {\nu_{ki}}{\epsilon^i_{\bar{b}}-\epsilon^k_{\bar{b}}} \qquad \qquad \qquad \    , \; \epsilon^i_{\bar{b}}\neq \epsilon^k_{\bar{b}} \nonumber
\\ c_{ki}&= &\frac{1}{\nu_{ii}-\nu_{kk}}\sum_{j\notin D_{k}}\frac {\nu_{kj}\nu_{ji}}{\epsilon^j_{\bar{b}}-\epsilon^k_{\bar{b}}}, \; \epsilon^i_{\bar{b}}= \epsilon^k_{\bar{b}}, \nu_{ii}\neq \nu_{kk} \nonumber
\\ c_{ki}&=&0 \qquad \qquad \qquad \qquad \quad  , \; \epsilon^i_{\bar{b}}= \epsilon^k_{\bar{b}}, \nu_{ii}= \nu_{kk}. \quad
\end{eqnarray}
 
I warn the reader that the equations~(\ref{eq:22}) and~(\ref{eq:23}) are \emph{exact} to the first order in~$\lambda$, for instance,
\begin{displaymath}
\lim_{\lambda \to 0} \frac{\epsilon^k_{\bar{b}, \lambda} - \epsilon^k_{\bar{b}}}{\lambda} = {\left\langle{\phi_{\bar{b}}^k} |V-V'_{\bar{b}}| {\phi_{\bar{b}}^k}\right\rangle},
\end{displaymath}

With these preparations, we are ready to prove an important lemma:
\begin{2}
\begin{eqnarray}
\label{eq:25}
&&\beta \int_{\Omega} \ud P'_{(x,x',\bar{b},\beta)}[\bar{a}] [ U(x,x',\bar{a};\beta) - U'_{\bar{b}}(x,x',\bar{a};\beta)] \nonumber 
 \\ &=& \frac{ \sum_{k =0}^{\infty} \sum_{j\neq k}^{\infty} c_{kj} \left[\phi_{\bar{b}}^k(x) {\phi_{\bar{b}}^{j}(x')}+\phi_{\bar{b}}^k(x') {\phi_{\bar{b}}^{j}(x)} \right] e^{-\beta \epsilon^k_{\bar{b}}} }
 {\sum_{k=0}^{\infty} \phi_{\bar{b}}^{k}(x) \phi_{\bar{b}}^{k}(x') e^{-\beta \epsilon^{k}_{\bar{b}}} } \nonumber 
 \\&& + \beta \frac{\sum_{k=0}^{\infty} \phi_{\bar{b}}^{k}(x) \phi_{\bar{b}}^{k}(x')  \langle \phi_{\bar{b}}^{k}|V-V'_{\bar{b}}| \phi_{\bar{b}}^{k}\rangle e^{-\beta \epsilon^{k}_{\bar{b}}}} {\sum_{k=0}^{\infty} \phi_{\bar{b}}^{k}(x) \phi_{\bar{b}}^{k}(x') e^{-\beta \epsilon^{k}_{\bar{b}}}}.
\end{eqnarray}
\end{2}

\emph{Proof:} 
Write $V''_{\bar{b}}(x)=V(x)-V'_{\bar{b}}(x)$ and 
\[U''_{\bar{b}}(x,x',\bar{a};\beta)=U(x,x',\bar{a};\beta)-U'_{\bar{b}}(x,x',\bar{a};\beta),\] 
so that: \[U''_{\bar{b}}(x,x',\bar{a};\beta)=\int_{0}^{1}V''_{\bar{b}}\big[x(t)+\sum_{k=1}^{\infty}a_k \sigma_k \sin(k \pi t) \big] \ud t.\]
Then consider the equality:
\begin{displaymath}
  \beta U''_{\bar{b}}(x,x',\bar{a};\beta)
 =\lim_{\lambda \rightarrow 0}\frac{1-\exp{\left[ -\lambda \beta U''_{\bar{b}}(x,x',\bar{a};\beta) \right]}}{\lambda}.
\end{displaymath}
Remembering the definition~(\ref{eq:14}) of the probability measure $P'_{(x,x',\bar{b};\beta)}[\bar{a}]$ and the eigenfunction series representation of a density matrix, we learn that 
\begin{widetext}
\begin{equation}
\label{eq:26}
 \beta \int_{\Omega}\, \ud P'_{(x,x',\bar{b};\beta)}[\bar{a}]U''_{\bar{b}}(x,x',\bar{a};\beta) =\lim_{\lambda \rightarrow 0} \frac{1}{\lambda} \frac{\sum_{k=0}^{\infty} \phi_{\bar{b}}^k (x) \phi_{\bar{b}}^k (x') e^{  -\beta \epsilon^k_{\bar{b}} } - \sum_{k=0}^{\infty} \phi_{\bar{b}, \lambda}^k (x) \phi_{\bar{b},\lambda}^k (x') e^{ -\beta \epsilon^k_{\bar{b}, \lambda} }}{ \sum_{k=0}^{\infty} \phi_{\bar{b}}^k (x) \phi_{\bar{b}}^k (x') e^{ -\beta \epsilon^k_{\bar{b}} }}, 
\end{equation}
\end{widetext}
where $\phi_{\bar{b}, \lambda}^k (z)$ and $ \epsilon^k_{\bar{b}, \lambda}$ are the eigenfunctions, respectively the eigenvalues of the perturbed Hamiltonian~(\ref{eq:21}).

For small $\lambda$, we may use the Rayleigh-Schr{\"o}dinger perturbation theory to compute the spectrum of the perturbed Hamiltonian $\hat{H}'_{\bar{b},\lambda}$. The reader should realize that the only corrections needed are the ones to the first order in $\lambda$ (which are \emph{exactly} given by the Rayleigh-Schr{\"o}dinger perturbation theory), since the others will cancel upon letting $\lambda$ go to zero. To conclude the proof, use the formulae~(\ref{eq:22}-\ref{eq:23}) and explicitly compute the limit in~(\ref{eq:26}).~$\Box$

Lemma 1 together with the well-known series representation of a density matrix 
\begin{equation}
\label{eq:27}
\rho'_{\bar{b}}(x,x';\beta)=\sum_{k=0}^{\infty} \phi_{\bar{b}}^{k}(x) \phi_{\bar{b}}^{k}(x')  e^{-\beta \epsilon^{k}_{\bar{b}}}
\end{equation}
essentially solves the eigenfunction representation problem. The functions
\begin{equation}
\label{eq:28}
T_{\bar{b}}^{(1)}(x,x';\beta)= \sum_{j\neq k}\! c_{kj} \! \left[\phi_{\bar{b}}^k(x) {\phi_{\bar{b}}^{j}(x')}\!+ \phi_{\bar{b}}^k(x') {\phi_{\bar{b}}^{j}(x)} \right]\! e^{-\beta \epsilon^k_{\bar{b}}}
\end{equation}
and
\begin{equation}
\label{eq:29}
T_{\bar{b}}^{(2)}(x,x';\beta)=\sum_{k=0}^{\infty} \phi_{\bar{b}}^{k}(x) \phi_{\bar{b}}^{k}(x')  \langle \phi_{\bar{b}}^{k}|V-V'_{\bar{b}}| \phi_{\bar{b}}^{k}\rangle e^{-\beta \epsilon^{k}_{\bar{b}}}
\end{equation}
have the following obvious properties:
\begin{equation}
\label{eq:30}
\int_{\mathbb{R}}T_{\bar{b}}^{(1)}(x;\beta) \ud x=0
\end{equation}
and respectively,
\begin{eqnarray}
\label{eq:31}
\int_{\mathbb{R}}T_{\bar{b}}^{(2)}(x;\beta) \ud x&=&\sum_{k=0}^{\infty}  \langle \phi_{\bar{b}}^{k}|V-V'_{\bar{b}}| \phi_{\bar{b}}^{k}\rangle e^{-\beta \epsilon^{k}_{\bar{b}}}
\nonumber \\ &=& Tr\left[(\hat{H}-\hat{H}'_{\bar{b}})e^{-\beta \hat{H}'_{\bar{b}}}\right].
\end{eqnarray}
Therefore,
\begin{3} The eigenfunction expansion form of the local variational inequality~(\ref{eq:20}) is
\begin{eqnarray}
\label{eq:32}
\rho(x,x';\beta) &\geq& \rho_{\bar{b}}^{a}(x,x';\beta)=\rho'_{\bar{b}}(x,x';\beta)  \nonumber \\&\times& \! \exp\! \left[-\frac{T_{\bar{b}}^{(1)}(x,x';\beta)\!+\beta T_{\bar{b}}^{(2)}(x,x';\beta)}{\rho'_{\bar{b}}(x,x';\beta)}\right]. \qquad
\end{eqnarray}\end{3} 

Before continuing, the reader is advised to ponder over the value of this theorem by analyzing the HO-LVP approximation $\rho^a_{z,\omega}(x,x';\beta)$ compactly given as a \emph{monodimensional} integral against~$t$ by Eq.~\ref{eq:20aa}. The same HO-LVP approximation can be exactly written in terms of the eigenvalues
\[\epsilon^{k}_{z,\omega}=\hbar \omega \left(k+\frac{1}{2}\right)\]
and the eigenfunctions
\[
\phi^{k}_{z,\omega}(x)=(2^k k!)^{-1/2}\left(\frac{m_0 \omega} {\pi \hbar} \right)^{1/4}e^{\zeta^2/2}H_{k}(\zeta)\]
 of the harmonic oscillator reference potential~(\ref{eq:20a}), in the form given by Theorem~2.  Here, $\zeta=(m_0\omega/\hbar)^{1/2}(x-z)$, while $H_k(x)$ stands for the respective Hermite polynomial. Of course, the eigenfunction representation is of no practical use, but it allows us to study the low temperature behavior of the density matrix.  

In the form~(\ref{eq:32}), the local variational inequality is readily seen to imply the Gibbs-Bogoliubov inequality. Indeed, setting $x=x'$, integrating over $x$ and applying Jensen's inequality produces:
\begin{eqnarray}
\label{eq:33}
e^{-\beta F}&\geq& e^{-\beta F'_{\bar{b}}}\int_{\mathbb{R}}\ud x\frac{\rho'_{\bar{b}} (x;\beta)} {e^{-\beta F'_{\bar{b}}}} \nonumber \\ && \times \exp \left[-\frac{T_{\bar{b}}^{(1)}(x;\beta)+\beta T_{\bar{b}}^{(2)}(x;\beta)}{\rho'_{\bar{b}}(x;\beta)}\right] \nonumber \\ & \geq &
e^{-\beta F'_{\bar{b}}}\exp\left\{-\beta \frac{Tr\left[(\hat{H}-\hat{H}'_{\bar{b}})e^{-\beta \hat{H}'_{\bar{b}}}\right]}{Tr(e^{-\beta \hat{H}'_{\bar{b}}})}\right\}, \qquad
\end{eqnarray}
where we used the relations~(\ref{eq:30}) and~(\ref{eq:31}). The last equation produces the Gibbs-Bogoliubov inequality upon taking the logarithm. Moreover, since we performed the same operations as for~(\ref{eq:18}), we also get a proof of the equivalence between Feynman and Gibbs-Bogoliubov inequalities, provided that the form~(\ref{eq:19}) for the trial potential is assumed. It is in this respect that we regard LVP as a generalization of both aforementioned inequalities, even if the best density matrix predicted is not necessarily derivable from a potential.

The remainder of this section deals with the low temperature behavior of the LVP density matrix $\rho_{best}(x,x';\beta)$.
An immediate corollary of Lemma 1 is the equality
\begin{eqnarray}
\label{eq:34}
&\lim_{\beta \rightarrow \infty}&\beta \bigg\{ \int_{\Omega}\, \ud P'_{(x,x',\bar{b};\beta)}[\bar{a}] \Big[U(x,x',\bar{a};\beta) \nonumber \\  &&-U'_{\bar{b}}(x,x',\bar{a};\beta)\Big]- {\left\langle {\phi_{\bar{b}}^0} |V - V'_{\bar{b}}| {\phi_{\bar{b}}^0} \right\rangle} \bigg\} 
\nonumber \\&&= S_{\bar{b}}(x)+S_{\bar{b}}(x'),
\end{eqnarray}
where
\begin{equation}
\label{eq:35} 
S_{\bar{b}}(x)= \sum_{k=1}^{\infty} \frac{\phi_{\bar{b}}^k(x)} {{\phi_{\bar{b}}^0(x)}} \frac{\left\langle {\phi_{\bar{b}}^0} |V - V'_{\bar{b}}| {\phi_{\bar{b}}^k} \right\rangle}{\epsilon^k_{\bar{b}} - \epsilon^0_{\bar{b}}}
\end{equation} 
is a function which does not depend upon temperature. In deducing~(\ref{eq:35}), one  uses the fact that the groundstate eigenfunction of the trial potential~$V'_{\bar{b}}(x)$ is not degenerate.  Then, the asymptotic formula \[ \rho'_{\bar{b}}(x,x'; \beta) \approx \phi^0_{\bar{b}}(x) \phi^0_{\bar{b}}(x') \exp(-\beta \epsilon^0_{\bar{b}}) \] implies
\begin{eqnarray}
\label{eq:36}
\rho_{\bar{b}}^{a}(x,x';\beta) &\approx&   \phi^0_{\bar{b}} (x) \phi^0_{\bar{b}} (x') \exp{ \left\{ -\left[S_{\bar{b}}(x)+S_{\bar{b}}(x')\right] \right\} }  \nonumber \\&& \times \exp \left[-\beta \left( \epsilon^0_{\bar{b}} + {\left\langle {\phi_{\bar{b}}^0} |V - V'_{\bar{b}}| {\phi_{\bar{b}}^0} \right\rangle} \right) \right]. \quad
\end{eqnarray}
I warn the reader that here and in the remainder of the paper the sign $\approx$ is used to denote a \emph{low temperature asymptotic form}, its \emph{rigorous} interpretation being:
\begin{eqnarray*}
\lim_{\beta \rightarrow \infty} \left\{ \exp \left[\beta \left( \epsilon^0_{\bar{b}} + {\left\langle {\phi_{\bar{b}}^0} |V - V'_{\bar{b}}| {\phi_{\bar{b}}^0} \right\rangle} \right) \right] \rho_{\bar{b}}^{a}(x,x';\beta)\right\}=\\= \phi^0_{\bar{b}} (x) \phi^0_{\bar{b}} (x') \exp{ \left\{ -\left[S_{\bar{b}}(x)+S_{\bar{b}}(x')\right] \right\} }.
\end{eqnarray*}

Looking at the equation~(\ref{eq:36}), we see that the factor containing the $S_{\bar{b}}(x)$ functions simply disappears in the original GBF equation because of the identity~(\ref{eq:30}). Thus, our theory brings some new information about the shape of the groundstate density matrix, and we shall later prove that the correction factor is always an improvement in the energetic sense. After an obvious simplification of the terms explicitly involving the potential~$V'_{\bar{b}}(x)$, the following theorem is immediate:
\begin{4}
The asymptotic formula of~$\rho_{best}(x,x';\beta)$ at low temperature is:
\begin{eqnarray}
\label{eq:37}
\rho_{best}(x,x';\beta)& \approx& \phi^0_{\bar{b}} (x) \phi^0_{\bar{b}} (x') \exp{ \left\{ -\left[S_{\bar{b}}(x)+S_{\bar{b}}(x')\right] \right\} } \nonumber 
\\ &&\times \exp \left[-\beta E( \phi^0_{\bar{b}} )\right] \Big{|}_{\bar{b} =\bar{B}(x,x',\infty)},
\end{eqnarray}
where
\begin{equation}
\label{eq:38}
E(\psi) = \int_{\mathbb{R}}\left[ \frac{\hbar^2}{2m} \|\nabla{\psi}(x) \|^2 + \psi(x)^2V(x) \right] \ud x 
\end{equation}
and the functions $\bar{B}(x,x', \infty)$ are computed by the following recipe:
\flushleft
\begin{enumerate}
\item
Minimize the functional $E(\phi^0_{\bar{b}})$. If it is unique, the value of $\bar{b}$ on which the minimum is attained becomes $\bar{B}(x,x', \infty) \; \forall \, x,x'\in \mathbb{R}$.
\item
If there are multiple minima of $E(\phi^0_{\bar{b}})$, pick an arbitrary one among those that further maximizes \[ \phi^0_{\bar{b}} (x) \phi^0_{\bar{b}} (x') \exp{ \Big\{ -\big[S_{\bar{b}}(x)+S_{\bar{b}}(x')\big] \Big\} } \qquad \qquad \] 
 at each pair of points $(x,x')$. 
\end{enumerate} 
\end{4}

Let us analyze a little more closely what the last theorem says. Assume we are in the simple case when the minimum of the functional $E(\phi^0_{\bar{b}})$ is unique. Up to a normalization factor, Theorem~3 predicts the following approximation to the groundstate eigenfunction:
\begin{equation}
\label{eq:38a}
\psi_{\bar{b}}(x)= \phi^0_{\bar{b}}(x)\exp{\left\{- \sum_{k=1}^{\infty} \frac{\phi_{\bar{b}}^k(x)} {{\phi_{\bar{b}}^0(x)}} \frac{\left\langle {\phi_{\bar{b}}^0} |V - V'_{\bar{b}}| {\phi_{\bar{b}}^k} \right\rangle}{\epsilon^k_{\bar{b}} - \epsilon^0_{\bar{b}}}\right\}}, \end{equation} 
where the optimal parameters~$\bar{b}$ do not depend upon the coordinates $(x,x')$. Thus, the Theorem~3 does not simply predict the function~$\phi_{\bar{b}}^0(x)$, though the thermodynamic weight is computed with respect to this function. An immediate question is in place: What can we say about the quality of the above eigenfunction? The quite remarkable answer is proved in the next section (see Eq.~\ref{eq:73}), and says that the expected energy of $\psi_{\bar{b}}(x)$ is always smaller or equal to the expected energy of $\phi_{\bar{b}}^0(x)$. In other words, LVP predicts an energetically better groundstate eigenfunction, and we shall prove in Section~V that we can recover its expected energy by use of the H-estimator. Finally, for multidimensional systems,  LVP  predicts a \emph{correlated} approximation of the groundstate eigenfunction even if the reference is a sum of monoparticle potentials. The low temperature density matrix given by Eq.~\ref{eq:37} has no GBF equivalent and justifies our claim that LVP is a separate and more powerful principle. 

\section{The expected energy of the LVP groundstate density matrix}

Let us remember that our special interest for the groundstate density matrix is due to our experience that the various approximations used to compute the finite-temperature statistical properties of a physical system  worsen in the low temperature regime. By the intrinsic continuity of the variational methods (see Section~V for further clarifications), a good approximation of the groundstate density matrix necessarily implies a good approximation for the finite-temperature density matrix. In this section, we shall analyze the expected energy of the groundstate density matrix predicted by Theorem~3, but we assume a special form of the density matrix which is encountered in practical applications whenever the potential $V(x)$ has a finite number of local minima. 

There is one special parameter $b_{0}$ which accounts for a translation and which we add to the list of parameters $\bar{b}=(b_1,b_2, \ldots)$. From now on, we shall conform to the convention that if not written explicitly in an expression, $b_0$ is assumed to be part of the list of parameters $\bar{b}$ i.e., $\bar{b}=(b_0, b_1 \ldots)$. Otherwise, if $b_0$ does appear in an expression, the list $\bar{b}$ is assumed not to contain it. The importance of this parameter consists of the fact that, if it is included,  the optimizing coefficients $\bar{B}(x,x';\infty)$ usually become constant on certain regions of the configuration space, which are identified with the main wells of the potential. Of course, for an $n$-dimensional system, there are~$n$ translational parameters, one for each dimension. 

To begin with, we replace~(\ref{eq:19}) by the slightly more general form
\begin{equation}
\label{eq:39}
U'_{b_{0},\bar{b}}(x,x',\bar{a};\beta)=\int_{0}^{1}\!V'_{\bar{b}} \big[-b_0+ x(t)+ \sum_{k=1}^{\infty}a_k \sigma_k \sin(k \pi t) \big]\ud t.
\end{equation}
According to our convention, $V'_{\bar{b}}(x)$ does not depend explicitly upon $b_0$, the value of this parameter, which sets the origin of the potential, being automatically determined by LVP. All our results remain true if computed with respect to the local reference potential $V'_{b_{0},\bar{b}}(x)=V'_{\bar{b}}(x-b_{0})$. If the eigenfunctions of $V'_{b_{0},\bar{b}}(x)$ are~$\phi^k_{b_0,\bar{b}}(x)$ and the corresponding eigenvalues are~$\epsilon^k_{b_0,\bar{b}}$, then we have $\phi^k_{b_0,\bar{b}}(x)=\phi^k_{\bar{b}}(x-b_0)$ and $\epsilon^k_{b_0,\bar{b}}=\epsilon^k_{\bar{b}}$.  It is then convenient to introduce the two local quantities: 
\begin{equation}
\label{eq:40}
V_z(x)=V(x+z)
\end{equation}
and
\begin{equation}
\label{eq:41}
E_{z}(\psi) = \int_{\mathbb{R}}\left[ \frac{\hbar^2}{2m} \|\nabla{\psi}(x) \|^2 + \psi(x)^2V(x+z) \right] \ud x. 
\end{equation}
Also, we shall use $z$ instead of $b_0$ and let $\mathcal{A}$ index all the pairs $(z_\alpha, \bar{B}_\alpha)$ on which the minimum of the problem
\begin{equation}
\label{eq:42}
E_{best}=\inf_{z,\bar{b}} E_{z}(\phi_{\bar{b}}^0)
\end{equation}
is achieved.
The new system of indexation makes the old one superfluous, so we shall drop some indices. We define:
\begin{equation}
\label{eq:43}
S_{\alpha}(x)= \sum_{k=1}^{\infty} \frac{\phi_{\alpha}^k(x-z_\alpha)} {{\phi_{\alpha}^0(x-z_\alpha)}} \frac{\langle {\phi_{\alpha}^0} |V_{z_\alpha} - V'_{\bar{B}_\alpha}| {\phi_{\alpha}^k} \rangle}{\epsilon^k_{\alpha} - \epsilon^0_{\alpha}} 
\end{equation}
and
\begin{equation}
\label{eq:44}
\psi_{\alpha}(x)=\phi_{\alpha}^0(x-z_\alpha)\exp[-S_{\alpha}(x)],
\end{equation}
where $\phi_{\alpha}^k(x)$ and $\epsilon_{\alpha}^{k}$ are the eigenfunctions, respectively the eigenvalues of the trial potential $V'_{\bar{B}_{\alpha}}(x)$.

With these notations, Theorem~3 takes on the special form:
\begin{5}
If~(\ref{eq:39}) is assumed, then the asymptotic formula of~$\rho_{best}(x,x';\beta)$ at low temperature is:
\begin{equation}
\label{eq:45}
\rho_{best}(x,x';\beta) \approx \exp (-\beta E_{best}) \rho^{\circ}_{best}(x,x'),
\end{equation}
where $E_{best}$ is the defined by~(\ref{eq:42}) and
\begin{equation}
\label{eq:46}
\rho^{\circ}_{best}(x,x')= \sup_{\alpha \in \mathcal{A}} \psi_{\alpha}(x)\psi_{\alpha}(x'). 
\end{equation}
\end{5}

To appreciate the importance of the translational parameter $z$, let us perform the minimization in~(\ref{eq:42}) in two separate steps. Firstly, we construct an effective potential 
\begin{equation}
\label{eq:47}
V_{ef}(z)=\inf_{\bar{b}}E_z(\phi_{\bar{b}}^{0})
\end{equation}
and secondly, we compute
\begin{equation}
\label{eq:48} 
E_{best}=\inf_{z}V_{ef}(z).
\end{equation}
For monodimensional systems, it is usually the case that the minimum of the first problem is attained on unique points $\bar{b}\equiv \bar{B}(z)$ while for multidimensional ones (especially for systems in condensed phase) there is usually a finite number of minimizing parameters. The effective potential is in fact a mollification of the original potential $V(z)$, to which it converges as the ratio $\hbar^2/m$ goes to zero. For systems in condensed phase, it is often the case that both the original and the effective potentials have finitely many global minima, and in this section we shall assume that there are finitely many pairs $(z_i, \bar{B}_i), i\in\overline{1,N}$ on which $\inf_{z,\bar{b}} E_{z}( \phi^0_{\bar{b}})$ is attained. A more general result will be proved in Section~V. If we set 
\begin{equation}
\label{eq:49}
 D_i=\left\{ (x,x')\in \mathbb{R}^2:\rho^{\circ}_{best}(x,x')= \psi_{i} (x) \psi_{i}  (x') \right\},
\end{equation}
then the sets $D_i$ are assumed to be disjoint except for their (topological) frontiers which are required to have measure zero. It follows that the optimizing coefficients are constant on the interior of the sets $D_i$ and  that $\rho^{\circ}_{best}(x,x')$ is twice derivable with continuous derivatives on the same interiors, yet continuous on the entire plane $\mathbb{R}^2$. Therefore, the diagonal density matrix $\rho^{\circ}_{best}(x)$ as well as its square root have similar continuity properties with respect to the \emph{diagonal} sets $D_i^\pi$, defined as the intersections of the $D_i$'s with the line of equation $x=x'$ [remember the convention $\rho^{\circ}_{best}(x)\equiv \rho^{\circ}_{best}(x,x)$]. 
	
	\begin{figure}[tbp]  \centering
   \includegraphics[angle=270,width=8.5cm,clip=t]{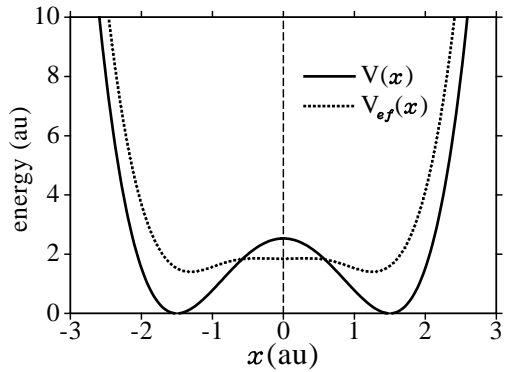} 
 \caption[sqr]
{\label{Fig:1}
A plot of the symmetric double well quartic potential~(\ref{eq:50}) and of its associated effective potential~(\ref{eq:47}).
}
\end{figure}
	
	The rest of this section deals with the evaluation of the expected energy of $\rho^{\circ}_{best}(x,x')$. To reinforce the proofs, we study  a simple example of a quartic double well potential in the context of the HO-LVP approximation, along with the general approach. We shall set $\hbar=1$, and consider a particle of mass $m=1$ moving in the potential
\begin{equation}
\label{eq:50}
V(x)=\frac{1}{2}(x-A)^2(x+A)^2,
\end{equation}
where $A=1.5$ (see Fig.~1). The reference potential is a quadratic one, of variable frequency $\omega>0$:
\begin{equation}
\label{eq:51}
V'_{\omega}(x)=\frac{1}{2}m\omega^2 x^2.
\end{equation} 
The functional~(\ref{eq:41}) can be worked out explicitly to be
\begin{equation}
\label{eq:52}
E_{z}(\omega)=\frac{\omega}{4}+\frac{3z^2-A^2}{2\omega}+\frac{3}{8\omega^2}+V(z).
\end{equation} 
Fig.~1 also contains a plot of the effective potential 
\[
V_{ef}(z)=\inf_{\omega>0}E_{z}(\omega)
\]
and shows that $V_{ef}(z)$ attains its global minimum $E_{best}=1.404$ on the two symmetric points $z_1=-1.292$ and $z_2=1.292$. The corresponding optimum reference potential frequencies are $\omega_{1}=\omega_{2}=\omega=2.584$, equal by the symmetry of the problem. The $S_{i}(x)$ functions can be expressed in terms of the Hermite polynomials as
\begin{equation}
\label{eq:53}
S_i(x)=\sum_{k=3,4}c_k^i \frac{1}{\sqrt{2^k k!}}\frac{H_k[\omega^{1/2} (x-z_i)]} {k\omega},
\end{equation}
where $c_3^1=-0.539$, $c_3^2=0.539$ and $c_4^1=c_4^2=0.092$. In general, it can be shown that all the coefficients $c_k$ with $k>2n$ vanish for any polynomial potential of rank at most $2n$, while the coefficients $c_1$ and $c_2$ vanish for all potentials. The functions $\psi_{i}(x)$  have the  form
\begin{equation}
\label{eq:54}
\psi_i(x)=\exp\left[-\frac{1}{2}{\omega(x-z_i)^2}  -S_i(x)\right],
\end{equation}
so that 
\begin{equation}
\label{eq:55}
\rho_{best}^{\circ}(x,x')=\max_{i\in\{1,2\}}\psi_i(x)\psi_i(x').
\end{equation}
In Fig.~2, one may see that the low temperature density matrix predicted by LVP is symmetrical at reflection with respect to both the main and the secondary axis. Though continuous on the entire plane, the density matrix has a cusp along the secondary axis.  The sets $D_i$ are readily identified: $D_1=\{(x,x'): x'<x\}$ and $D_2=\{(x,x'): x'>x\}$ with the diagonals $D_1^\pi=\{x<0\}$ and $D_2^\pi=\{x>0\}$. 
\begin{figure}[tbp] 
\centering
\includegraphics[angle=270,width=8.5cm,clip=t]{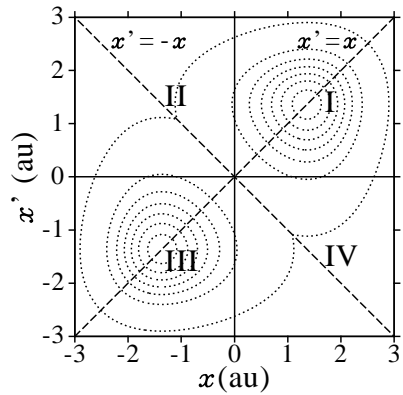} 
\caption[sqr] {\label{Fig:2}
A plot of the low temperature density matrix predicted by LVP.  There are only two maxima instead of four symmetrical ones, the true density matrix would present.}
\end{figure}

\begin{figure}[!b] 
\centering
\includegraphics[angle=270,width=8.5cm,clip=t]{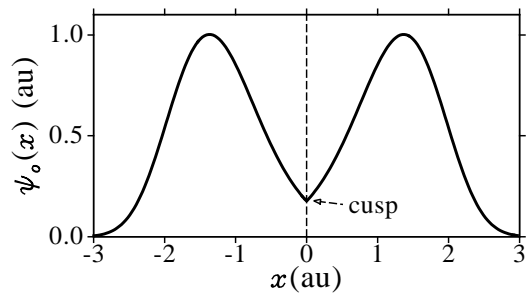} 
\caption[sqr]{\label{Fig:3}
For the double well quartic potential~(\ref{eq:50}), the approximate groundstate wavefunction $\psi_\circ(x)$ defined by~(\ref{eq:59}) has a cusp at the origin.}
\end{figure}

Let us go back to the energy evaluation problem. The lack of continuity of the first derivatives on the boundaries $\partial D_i$ requires a  careful analysis of the kinetic energy. We consider two estimators:
\begin{equation}
\label{eq:56}
K_{1}(\rho)=\frac{\hbar^2}{2m} \frac{\int_{\mathbb{R}}\frac{\partial^2}{\partial x \partial x'}\rho(x,x')\big|_{x=x'} \ud x}{\int_{\mathbb{R}} \rho(x) \ud x}
\end{equation}
and
\begin{equation}
\label{eq:57}
K_{2}(\rho)=-\frac{\hbar^2}{2m} \frac{\int_{\mathbb{R}}\frac{\partial^2}{\partial x^2}\rho(x,x')\big|_{x=x'} \ud x}{\int_{\mathbb{R}} \rho(x) \ud x},
\end{equation}
where the derivatives are regarded as functions almost everywhere defined and not as distributions. We shall denote by $E_1(\rho)$ and $E_2(\rho)$ the associated energy estimators, obtained by adding the expected potential energy. 

A little thought shows that the following equality holds for  all the points $(x,x')$ in the first and the third quadrants of the plane:
\begin{equation}
\label{eq:58}
\rho^{\circ}_{best}(x,x')=\sqrt{\rho^{\circ}_{best}(x)}\sqrt{\rho^{\circ}_{best}(x')}.
\end{equation}
In fact, the square root of the diagonal density $\rho^{\circ}_{best}(x)$, which has a cusp in the origin as shown in Fig.~3, will play  such an important role in our development that it deserves a notation:
\begin{equation}
\label{eq:59}
\psi_{\circ}(x)=\sqrt{\rho^{\circ}_{best}(x)}.
\end{equation}
In general, one may show that around each diagonal point $(x,x)$ on the interior of some set $D_i^\pi$, there is a small neighborhood in~$\mathbb{R}^2$, say the ball  $B[\epsilon,(x,x)]$, such that~(\ref{eq:58}) holds on $B[\epsilon,(x,x)]$. Relation~(\ref{eq:58}) needs not hold for the diagonal points $(x,x)$ that are precisely on some frontier $\partial  D_i^\pi$ but from the point of view of integration theory, this does not matter because the frontier has measure zero. 
Consequently, the following equalities are true:
\begin{equation}
\label{eq:60}
K_{1}(\rho^{\circ}_{best})
=\frac{\hbar^2}{2m} \frac{\int_{\mathbb{R}}\big\|\nabla  \psi_\circ(x)  \big\|^2 \ud x}{\int_{\mathbb{R}} \psi_\circ(x)^2 \ud x}
\end{equation}
and
\begin{equation}
\label{eq:61}
K_{2}(\rho^{\circ}_{best})=-\frac{\hbar^2}{2m} \frac{\int_{\mathbb{R}} \psi_\circ(x) \frac{\partial^2} {\partial x^2}\psi_\circ(x)\, \ud x}{\int_{\mathbb{R}} \psi_\circ(x)^2 \ud x}.
\end{equation}

Because $\rho_{best}^{\circ}(x)$ is continuous on the entire $\mathbb{R}$ by the way it was constructed, it can be proven that $\psi_\circ(x)$ is in the Sobolev space $H^{1,2}(\mathbb{R})$, thus a permissible trial function for the groundstate eigenfunction of the potential $V(x)$. For functions $\psi(x)$ in $H^{1,2}(\mathbb{R})$, the Rayleigh-Ritz principle states that  
\begin{equation}
\label{eq:62}
\epsilon_0 \leq \frac{\hbar^2}{2m} \frac{\int_{\mathbb{R}}\|\nabla  \psi (x) \|^2 \ud x}{\int_{\mathbb{R}} \psi(x)^2 \ud x}+ \frac{\int_{\mathbb{R}}V(x) \psi (x)^2  \ud x}{\int_{\mathbb{R}} \psi(x)^2 \ud x}.
\end{equation}

Consequently, the correct variational definition of the kinetic energy is given by the formula~(\ref{eq:56}) and we have our first important result:
\begin{equation}
\label{eq:63}
\epsilon_0 \leq E_1(\rho^{\circ}_{best}).
\end{equation} 
For the case of the quartic potential, the exact values are: $\epsilon_0=1.292$ and $E_1(\rho^{\circ}_{best})=1.342$. We see that our estimation of the groundstate energy is better than the one given by GBF, which is $E_{best}=1.404$. We shall prove that this is no mistake, and that the energy of the asymptotical low temperature density matrix predicted by LVP is always lower than the one predicted by the GBF. We do this in two steps: first we prove that $E_1(\rho^{\circ}_{best})\leq E_2(\rho^{\circ}_{best})$  and then that $E_2(\rho^{\circ}_{best})\leq E_{best}$.

Integration by parts produces
\begin{eqnarray}
\label{eq:64}
-\int_{\mathbb{R}} \psi_\circ(x)\frac{\partial^2} {\partial x^2}\psi_\circ(x) \ud x&=& 
 \frac{1}{2}\sum_{i=1}^{N-1}\frac{\partial \left[ \psi_{i+1}^2- \psi_{i}^2 \right] }{\partial x}(x_i) \nonumber \\ 
&+&\int_{\mathbb{R}}\big\|\nabla  \psi_\circ(x)  \big\|^2 \ud x,
\end{eqnarray} 
where the points $x_i$ are separating two consecutive sets $D_i^\pi$ and $D_{i+1}^\pi$ on which the diagonal density matrix takes the values $\psi_i(x)^2$ and $\psi_{i+1}(x)^2$ respectively.
Notice that $\psi_i(x_i)^2=\psi_{i+1}(x_i)^2$ and that \[\psi_{i+1}(x_{i}+h)^2 \geq\psi_{i}(x_i+h)^2\] for all positive and small enough $h$ or, otherwise, $\rho_{best}^{\circ}(x)=\psi_{i}(x)^2$ for some $x \in D_{i+1}^\pi$ contradicting the definition of the set. Therefore,
\begin{equation}
\label{eq:65}
\frac{\partial \! \left[\psi_{i+1}^2-\psi_{i}^2\right]}{\partial x}(x_i) = \lim_{h \searrow 0} \! \frac{\psi_{i+1}^2(x_i+h)-\psi_{i}^2(x_i+h)}{h}\geq 0,
\end{equation} 
which together with~(\ref{eq:64}) proves that
\begin{equation}
\label{eq:66}
E_1(\rho^{\circ}_{best})\leq E_2(\rho^{\circ}_{best}).
\end{equation}
For multidimensional systems, the same reasoning can be performed along the normals $\bar{\nu}_{ij}$ to the surfaces $\partial D_i^\pi \cap \partial D_j^\pi$ separating the sets $D_i^\pi$ and $D_j^\pi$. The normal $\bar{\nu}_{ij}$ is assumed oriented from the $D_i^\pi$ toward the $D_j^\pi$ set. Then, the analog of~(\ref{eq:65}) is
\begin{equation}
\label{eq:67}
\nabla \! \left[\psi_{j}^2-\psi_{i}^2\right]\cdot \bar{\nu}_{ij} \geq 0 \quad \text{on} \ \partial D_i^\pi \cap \partial D_j^\pi 
\end{equation}
and the analog of~(\ref{eq:64}) is
\begin{eqnarray}
\label{eq:68}
&-&\int_{\mathbb{R}^n}\psi_\circ(x)\Delta\psi_\circ(x)\ud x= 
\int_{\mathbb{R}^n}\big\|\nabla  \psi_\circ(x) \big\|^2 \ud x  \nonumber \\
&+&\frac{1}{2}\sum_{i<j}\int_{\partial D_i^\pi \cap \partial D_j^\pi }\!\! 
\nabla \! \left[\psi_{j}(x)^2-\psi_{i}(x)^2\right]\cdot \bar{\nu}_{ij} \ud \sigma, \quad
\end{eqnarray}
proving again~(\ref{eq:66}).

Finally, let us show that $E_2(\rho^{\circ}_{best}) \leq E_{best}$. Because $\phi^0_{i}$ is strictly positive, we can write any other eigenfunction $\phi^k_{i}(x)$ as the product $f^k_{i}(x)\phi^0_{i}(x)$. By direct substitution, one can show that the function $f^k_{i}(x)$ satisfies the following equation:
\begin{equation}
\label{eq:69}
-\frac{\hbar^2}{2m}\Delta f^k_{i}(x)-\frac{\hbar^2}{2m} \nabla \ln[\phi^0_{i}(x)^2] \cdot \nabla f^k_{i}(x)=(\epsilon^k_{i}-\epsilon^0_{i}) f^k_{i}(x).
\end{equation}
It follows then that
\begin{eqnarray}
\label{eq:70}
-\frac{\hbar^2}{2m}\Delta S_i(x)-\frac{\hbar^2}{2m} \nabla \ln[\phi^0_{i}(x-z_i)^2] \cdot \nabla S_i(x)\nonumber 
\\  = \sum_{k=1}^{\infty} f^k_{i}(x-z_i)\langle {\phi_{i}^0} |V_{z_i}\! - V'_{\bar{B}_i}| {\phi_{i}^k} \rangle. 
\end{eqnarray}
The sum of the last series  equals \[ V(x) - V'_{\bar{B}_i}(x-z_i)-\langle {\phi_{i}^0}  |V_{z_i}\! - V'_{\bar{B}_i}| {\phi_{i}^0} \rangle  \] by the completeness of the system of eigenfunctions $\{ \phi_{i}^k(x); k \geq 0\} $ and the translational invariance of the integrals involved, so we end up with the equality
\begin{eqnarray}
\label{eq:71}
-\frac{\hbar^2}{2m}\Delta S_i(x)-\frac{\hbar^2}{2m} \nabla \ln[\phi^0_{i}(x-z_i)^2] \cdot \nabla S_i(x) \nonumber 
\\ = [V(x) - V'_{\bar{B}_i}(x-z_i)] - \langle {\phi_{i}^0} |V_{z_i} - V'_{\bar{B}_i}|  {\phi_{i}^0} \rangle.  
\end{eqnarray}
With the help of~(\ref{eq:71}), one can show by explicit computation that
\begin{eqnarray}
\label{eq:72}
&&-\frac{\hbar^2}{2m}\Delta \psi_{i}(x) + V(x)\psi_{i}(x)  \nonumber 
\\&=& \Big[ \epsilon^0_{i} +\langle {\phi_{i}^0} |V_{z_i} - V'_{\bar{B}_i}|  {\phi_{i}^0} \rangle  -\frac{\hbar^2}{2m}\|\nabla S_i(x)\|^2 \Big] \psi_{i}(x) \nonumber
\\&=& \Big[ E_{best} -\frac{\hbar^2}{2m}\|\nabla S_i(x)\|^2 \Big] \psi_{i}(x).
\end{eqnarray}
We then multiply equation~(\ref{eq:71}) by $\psi_{i}(x)$, integrate over the sets $D_i^\pi$, sum the contributions of all sets and conclude that
\begin{equation}
\label{eq:73}
E_2(\rho_{best}^{\circ})= E_{best} -\frac{\hbar^2} {2m} \frac{\sum_{i=1}^{N}\int_{D_i^\pi}  \rho^{\circ}_{best}(x)\|\nabla S_i(x)\|^2 \ud x} {\int_{\mathbb{R}} \rho^{\circ}_{best}(x)\ud x}.
\end{equation}
Indeed, for the case of the quartic potential, one computes $E_2(\rho_{best}^\circ)=1.372$ which is seen to be lower than $E_{best}=1.404$ but higher than $E_1(\rho_{best}^\circ)=1.342$. 
The relations~(\ref{eq:63}), (\ref{eq:66}) and~(\ref{eq:73}) combined give
\begin{equation}
\label{eq:74}
\epsilon_0 \leq E_1(\rho_{best}^{\circ}) \leq E_2(\rho_{best}^{\circ}) \leq E_{best}, 
\end{equation}
which proves our previous assertion that the asymptotic density matrix predicted by LVP has a lower energy than the one given by GBF\@. In fact, if $E_{best}<\infty$, the last inequality is strict except for the case when the original potential and the optimized trial potential are identical.

\section{Average energy at low temperature. The Semi-Sum Theorem}
The LVP approximation is intended as a technique for computing finite-temperature properties of a quantum physical system, properties that are usually of the form
\[
\langle O \rangle_{\beta}=\frac{\int_{\mathbb{R}}\ud x \rho_{best}(x;\beta)O(x;\beta)} {\int_{\mathbb{R}} \ud x \rho_{best}(x;\beta)}.
\]
Such averages can be estimated for fairly complex systems by Monte Carlo simulations~\cite{Dol90}. The problem we address in this section is the low temperature limit of different energy estimators. For operators which are diagonal in the coordinate representation, for example the potential energy~$V(x)$, the estimating function $O(x)$ does not depend upon temperature and the zero temperature limit is
\[
\lim_{\beta \to \infty} \langle O \rangle_{\beta} = \frac{\int_{\mathbb{R}}\ud x \rho^{\circ}_{best}(x)O(x)} {\int_{\mathbb{R}} \ud x \rho^{\circ}_{best}(x)}.
\]

  In this paper, we assume that the pointwise optimization in the configuration space involved by LVP can be rapidly implemented by standard local optimization procedures, iterative methods or other approximations. Since this is a big assumption by itself, estimators explicitly depending upon  the derivatives of the optimizing parameters~$\bar{B}(x,x';\beta)$ are clearly out of question. In the remainder of this section, we shall consider the important problem of computing the ensemble average energy with the help of the so-called T, respectively  estimating functions, both of which can be put in a form that satisfies the aforementioned restriction. With regard to the zero temperature limit, we are interested to learn whether we can recover fully or only partially $E_1(\rho_{best}^\circ)$, the best energy predicted by LVP\@. We shall prove that we recover~$E_{best}$ with the help of the T-method estimator, respectively the semi-sum of~$E_1(\rho_{best}^\circ)$ and~$E_2(\rho_{best}^\circ)$ by the H-method estimator. The last fact is called the \emph{semi-sum theorem}.

We begin by considering some preliminary results. The maximum condition~(\ref{eq:20}) implies the equality
\begin{equation}
\label{eq:75}
\frac{\partial}{\partial \bar{b}}\rho_{\bar{b}}^a(x,x';\beta) \bigg|_{\bar{b}= \bar{B} (x,x';\beta)} =0 \qquad \forall \, x,x' \in \mathbb{R}.
\end{equation}
Another consequence of the same extremum condition is that the hessian matrix
\begin{equation}
\label{eq:76}
\frac{\partial^2}{\partial \bar{b}^2} \rho_{\bar{b}}^a(x,x';\beta) \bigg|_{\bar{b}= \bar{B} (x,x';\beta)}
\end{equation}
is negative definite $ \forall \, x,x' \in \mathbb{R}$. Moreover, the symmetry of $\rho_{\bar{b}}^a(x,x';\beta)$ in the arguments $x$ and $x'$ implies the symmetry of the minimizing functions $\bar{B}(x,x';\beta)$ in the same arguments. We then have the equality
\begin{equation}
\label{eq:77}
 \frac{\partial}{\partial x} \bar{B}(x,x';\beta) \bigg|_{x=x'}=\frac{\partial}{\partial x'} \bar{B}(x,x';\beta) \bigg|_{x=x'}.
\end{equation}
At finite temperature, because of the thermal averaging, it is safe to assume that the optimizing parameters $\bar{B}(x,x';\beta)$ are nice functions in their arguments with continuous partial derivatives at least to the first order. This might not be true for $\bar{B}(x,x';\infty)$ which may be constant on the interior of some sets~$D_i$, but vary suddenly at their frontier. 

For the rest of the paper, we shall assume that~$\rho_{best}^{\circ}(x,x')$ is in the Sobolev space~$H^{1,2}(\mathbb{R}^2)$. Thus, the norms (defined here by their square)
\begin{equation}
\label{eq:78}
\left\|\rho_{best}^{\circ}\right\|_0^2=\int_{\mathbb{R}}\ud x \! \! \int_{\mathbb{R}}\ud x' \rho_{best}^\circ(x,x')^2
\end{equation}
and
\begin{equation}
\label{eq:79}
\left\|\rho_{best}^{\circ}\right\|_1^2=\left\|\rho_{best}^{\circ}\right\|_0^2+ \int\! \! \int_{\mathbb{R}^2}\!\!\left[({\partial_x}\rho_{best}^\circ)^2+({\partial_{x'}}\rho_{best}^\circ)^2\right]
\end{equation}
are finite (for the case analyzed in the previous section, it is rather trivial to prove that these conditions are fulfilled). We shall also assume the existence of the second derivatives almost everywhere. Some mathematical difficulties force us to restrict the analysis to  potentials bounded from below---thus, positive by a change of reference. 

To avoid the excessive use of big vertical lines, we shall follow the rule that all functions $f(x,x',\bar{b};\beta)$ explicitly depending upon~$\bar{b}$ are evaluated at $\bar{b}=\bar{B}(x,x';\beta)$ if the results are to hold. The evaluation is done after any differentiation but before integration.

There are a couple of energy estimators in literature~\cite{Dol90}, of which we shall consider the most important two: the so called $T$-method and $H$-method estimators. The first one is computed by temperature differentiation of the canonical partition function
\begin{equation}
\label{eq:80}
\left\langle E \right\rangle^{T}_{\beta} =-\frac{\partial}{\partial \beta} \ln{\left[ \int_{\mathbb{R}} \rho_{best}(x;\beta) \ud x \right]}. 
\end{equation}
With the help of~(\ref{eq:75}), one can show that
\begin{equation}
\label{eq:81}
 \left\langle E \right\rangle^{T}_{\beta}  =-\frac{\int_{\mathbb{R}}  \frac{\partial}{\partial \beta} \rho_{\bar{b}}^a(x;\beta)\ud x} {\int_{\mathbb{R}}  \rho_{\bar{b}}^a(x;\beta) \ud x},
\end{equation}
expression that is seemingly easier to compute since it does not involve the evaluation of the partial derivatives of~$\bar{B}(x,x';\beta)$ with respect to temperature.
The low temperature limit is computed by replacing in formula~(\ref{eq:80}) the asymptotic density matrix given by~(\ref{eq:45}), to produce
\begin{equation}
\label{eq:82}
\lim_{\beta \rightarrow \infty}\left\langle E \right\rangle^{T}_{\beta}=E_{best}
\end{equation}
i.e., the groundstate energy we get by using the T-method estimator coincides with the best energy provided by the analog centroid based approximations.

In the particular case of the HO-LVP approximation, the diagonal density matrix takes the form:
\begin{eqnarray}
\label{eq:82a}
\rho^a_{z,\omega}(x;\beta)=\sqrt{\frac{m}{2\pi \hbar^2 \beta}}h_0(\beta C) \nonumber \quad \\ \times \exp{\Big[-\frac{1}{2}\beta^3B^2h_5(\beta C) +\frac{1}{2}\beta^2C^2h_4(\beta C)\Big]} 
\nonumber \quad \\ \times
 \exp{ \Bigg\{ -\beta\int_{0}^{1}\!\! \ud t \; \overline{V}_{t,\omega} \Big[x-}  
 \sigma\beta^{3/2}B h_2(\beta C,t) \Big] \Bigg\}, \quad
\end{eqnarray} 
where 
\begin{equation}
\label{eq:82b}
\sigma^2 = \frac{2 \beta \hbar^2}{\pi^2 m}.
\end{equation}
The T-estimator function has the expression
\begin{eqnarray}
\label{eq:82c}
E_{z,\omega}^{T}(x;\beta)= \frac{1}{2\beta}+ \int_{0}^{1}\!\! \ud t \; \overline{V}_{t,\omega} \Big[x- \sigma \beta^{3/2}B h_2(\beta C,t) \Big] \nonumber \\ - \frac{1}{2}\sigma \beta^{3/2} B\!\int_{0}^{1}\!\! \ud t \; \overline{V}'_{t,\omega} \Big[x- \sigma \beta^{3/2}B h_2(\beta C,t) \Big]h_{2}(\beta C,t) \nonumber \\+ \frac{1}{2}\sigma^2 \! \int_{0}^{1}\!\! \ud t \; \overline{V}''_{t,\omega} \Big[x- \sigma \beta^{3/2}B h_2(\beta C,t) \Big]h_{1}(\beta C,t) \quad
\end{eqnarray}

The $H$-method estimator is the direct expected value of the Hamiltonian
\begin{equation}
\label{eq:83}
\left\langle E \right\rangle^{H}_{\beta} =\frac{\int_{\mathbb{R}}\ud x \hat{H} \rho_{best}(x,x';\beta) \big|_{x=x'}} {\int_{\mathbb{R}}\ud x \rho_{best}(x;\beta)}.
\end{equation}
In computing the kinetic term of~(\ref{eq:83}), the following formula proves beneficial:
\begin{equation}
\label{eq:84}
\left\langle K \right\rangle^{H}_{\beta} =\frac{\hbar^2}{4m} \frac{\int_{\mathbb{R}}\ud x  \Big(\frac{\partial^2}{\partial x \partial x'}-\frac{\partial^2}{\partial x^2}\Big) \rho_{best}(x,x';\beta) \Big|_{x=x'}} {\int_{\mathbb{R}}\ud x \rho_{best}(x;\beta)}.
\end{equation}
We compute the expected kinetic energy as the semi-sum of the two \emph{identical} terms, because  this way no derivatives of $\bar{B} (x,x';\beta)$ appear in the final formula. By differentiation of~(\ref{eq:75}) against $x'$, we get the system of equations
\begin{equation}
\label{eq:85}
\frac{\partial^2}{\partial \bar{b} \partial x'} \rho_{\bar{b}}^a(x,x';\beta) + \frac{\partial^2}{\partial \bar{b}^2} \rho_{\bar{b}}^a(x,x';\beta) \frac{\partial \bar{B}(x,x';\beta)}{\partial x'}=0 
\end{equation}
and there is a similar one for the derivatives against $x$. From~(\ref{eq:75}) and~(\ref{eq:85}), the following equalities can be deduced by explicit calculation:
\begin{eqnarray}
\label{eq:86}
 &&\frac{\partial^2}{\partial x \partial x'} \rho_{best} (x,x';\beta )=  
\frac{\partial^2}{\partial x \partial x'} \rho_{\bar{b}}^a(x,x';\beta) \nonumber \\
&&-\frac{\partial^2}{\partial \bar{b}^2} \rho_{\bar{b}}^a(x,x';\beta) \frac{\partial \bar{B}(x,x';\beta)}{\partial x} \frac{\partial \bar{B}(x,x';\beta)}{\partial x'} \quad 
\end{eqnarray}
and 
\begin{eqnarray}
\label{eq:87}
 &&\frac{\partial^2}{\partial x^2} \rho_{best}(x,x';\beta )=  
\frac{\partial^2}{\partial x^2 } \rho_{\bar{b}}^a(x,x';\beta) \nonumber \\
&&-\frac{\partial^2}{\partial \bar{b}^2} \rho_{\bar{b}}^a(x,x';\beta) \frac{\partial \bar{B}(x,x';\beta)}{\partial x} \frac{\partial \bar{B}(x,x';\beta)}{\partial x}. \quad 
\end{eqnarray}
By adding~(\ref{eq:86}) and~(\ref{eq:87}) and using~(\ref{eq:77}), we get the equality
\begin{eqnarray}
\label{eq:88}
&&\left\{ \frac{\partial^2}{\partial x^2 } \rho_{best} (x,x';\beta )- \frac{\partial^2}{\partial x \partial x'} \rho_{best} (x,x';\beta )\right\}\bigg{|}_{x=x'} \nonumber \\&=& \left\{ \frac{\partial^2}{\partial x^2} \rho_{\bar{b}}^a(x,x';\beta)
- \frac{\partial^2}{\partial x \partial x'} \rho_{\bar{b}}^a(x,x';\beta) \right\}\bigg{|}_{x=x'}.
\end{eqnarray}
Relation~(\ref{eq:88}) shows that there is no need for the partial derivatives of the optimizing parameters~$\bar{B}(x,x';\beta)$ against $x$ or $x'$ in order to evaluate the ensemble average energy by the H-method estimator. We shall introduce two additional kinetic energy estimators which serve as intermediate tools in our computation: 
\begin{equation}
\label{eq:89}
 \left\langle K \right\rangle^{H,1}_{\beta} =\frac{\hbar^2}{2m} \frac{\int_{\mathbb{R}}\ud x  \frac{\partial^2}{\partial x \partial x'} \rho_{\bar{b}}^{a}(x,x';\beta) \Big|_{x=x'}} {\int_{\mathbb{R}}\ud x \rho_{best}(x;\beta)} 
\end{equation}
and
\begin{equation}
\label{eq:90}
 \left\langle K \right\rangle^{H,2}_{\beta} =-\frac{\hbar^2}{2m} \frac{\int_{\mathbb{R}}\ud x  \frac{\partial^2}{\partial x^2} \rho_{\bar{b}}^{a}(x,x';\beta) \Big|_{x=x'}} {\int_{\mathbb{R}}\ud x \rho_{best}(x;\beta)},
\end{equation}
and denote the respective energy estimators, obtained by adding the ensemble average potential energy, by $\langle E \rangle^{H,1}_{\beta}$ and $\langle E \rangle^{H,2}_{\beta}$ respectively. The second estimator, called in this work of ``type 2,'' is always greater than the first, which is called of ``type 1.'' Indeed, from~(\ref{eq:86}) and~(\ref{eq:87}) one learns that
\begin{eqnarray}
\label{eq:91}
&& \left\langle K \right\rangle^{H,2}_{\beta}-\left\langle K \right\rangle^{H,1}_{\beta}=\left\langle E \right\rangle^{H,2}_{\beta}- \left\langle E \right\rangle^{H,1}_{\beta} \nonumber \\
&&=-\frac{\hbar^2}{m} \frac{\int_{\mathbb{R}}\ud x  \frac{\partial^2}{\partial \bar{b}^2} \rho_{\bar{b}}^{a}(x;\beta)\frac{\partial \bar{B}(x,x';\beta)}{\partial x} \frac{\partial \bar{B}(x,x';\beta)}{\partial x} \Big|_{x=x'}} {\int_{\mathbb{R}}\ud x \rho_{best}(x;\beta)} \geq 0, \qquad
\end{eqnarray}
the inequality following from the negative definitiveness of the hessian matrix~(\ref{eq:76}).
Now, relation~(\ref{eq:88}) says that 
\begin{equation}
\label{eq:92}
 \left\langle K \right\rangle^{H}_{\beta} =\frac{1}{2}\left[\left\langle K \right\rangle^{H,1}_{\beta}+\left\langle K \right\rangle^{H,2}_{\beta}\right]
\end{equation} 
and so, 
\begin{equation}
\label{eq:93}
\left\langle E \right\rangle^{H}_{\beta} =\frac{1}{2}\left[\left\langle E \right\rangle^{H,1}_{\beta}+\left\langle E \right\rangle^{H,2}_{\beta}\right].
\end{equation} 

The reader might have already realized that the formula~(\ref{eq:93}) is the key to the semi-sum theorem announced at the beginning of the section. It also implies that there is no need for the partial derivatives of the optimizing parameters, in order to compute the H-method estimator. For the case of the HO-LVP, the H-estimator function has the expression
\begin{eqnarray}
\label{eq:93a}
E_{z,\omega}^{H}(x;\beta)= \frac{1}{2\beta}+ V(x)+ \beta^3 C^4 h_6(\beta C) \nonumber \\+ \frac{\pi^2\sigma^2}{8} \! \int_{0}^{1}\!\! \ud t \; \overline{V}''_{t,\omega} \Big[x- \sigma \beta^{3/2}B h_2(\beta C,t) \Big]h_7(\beta C, t), \quad
\end{eqnarray}
where~$\sigma$ is defined by Eq.~\ref{eq:82b}.

To continue, we turn our attention to some convergence problems. The expected values of the potential energy or other diagonal operators (including the constant functions) converge smoothly to the expected values computed with respect to~$\rho^{\circ}_{best}(x)$ as $\beta \rightarrow \infty$, fact that can be justified in most cases with the help of the Dominated Convergence Theorem (see Theorem~2.24 of Ref.~\onlinecite{Fol99}). However, this theorem cannot be used directly in the case of the expected values of the operators whose estimators explicitly involve the partial derivatives of~$\bar{B}(x,x';\beta)$. As we saw in the previous section, their moduli may blow up on~$\partial D_i$ so no dominating function might exist.

Using the asymptotic form~(\ref{eq:37}) predicted by Theorem~3, one easily proves the following analog of~(\ref{eq:73}): 
\begin{equation}
\label{eq:94}
\left\langle E \right\rangle^{H,2}_{\beta}\approx E_{best} -\frac{\hbar^2} {2m} \frac{\int_{\mathbb{R}}  \psi_{\bar{b}}(x)^2\|\nabla S_{\bar{b}}(x)\|^2  \ud x} {\int_{\mathbb{R}} \psi_{\bar{b}}(x)^2\ud x}.
\end{equation}
Since the expression on the right-hand side does not explicitly involve derivatives of the optimizing coefficients, we have the equality
\begin{equation}
\label{eq:95}
\lim_{\beta \rightarrow \infty}\left\langle E \right\rangle^{H,2}_{\beta}= E_{best}  - \frac{\hbar^2} {2m} \frac{\int_{\mathbb{R}}  \rho_{best}^\circ(x)\|\nabla S_{\bar{b}}(x)\|^2  \ud x} {\int_{\mathbb{R}} \rho_{best}^\circ(x)\ud x},
\end{equation}
where we set~${\bar{b}=\bar{B}(x,\infty)}$ before integration.
With the hypothesis that the potential is positive in mind, we see that~$\langle E \rangle^{H,2}_{\beta}$ is the biggest estimator around, so all above defined estimators are bounded by~$E_{best}$ for low enough temperature. Moreover, the estimator ~$\langle K \rangle^{H,1}_{\beta}$ cannot be negative since
\begin{equation}
\label{eq:96}
\lim_{\beta \rightarrow \infty}\left\langle K \right\rangle^{H,1}_{\beta}=  \frac{\hbar^2} {2m} \frac{\int_{\mathbb{R}}  \|\nabla \psi_{\bar{b}}(x)\|^2 \big|_{\bar{b}=\bar{B}(x,\infty)} \ud x} {\int_{\mathbb{R}} \rho_{best}^\circ(x)\ud x} \geq 0.
\end{equation}
It follows that the first term from the relation~(\ref{eq:91}) is bounded by~$E_{best}$ when~$\beta$ is large and again with the help of Theorem~3,  one may establish the result
\begin{widetext}\begin{eqnarray}
\label{eq:97}
\frac{m}{\hbar^2}E_{best}\int_{\mathbb{R}}\rho_{best}^\circ(x) \ud x &\geq& \lim_{\beta \rightarrow \infty} e^{\beta E_{best}}\int_{\mathbb{R}}\ud x  \frac{\partial^2 }{\partial \bar{b}^2}\rho_{\bar{b}}^{a}(x;\beta) \frac{\partial \bar{B}(x,x';\beta)}{\partial x} \frac{\partial \bar{B}(x,x';\beta)}{\partial x} \Big|_{x=x'}   \nonumber \\&=& M_1 \equiv \lim_{\beta \rightarrow \infty} \int_{\mathbb{R}}\ud x  \bigg[\beta\frac{\partial^2 E(\phi_{\bar{b}}^0)}{\partial \bar{b}^2}-\frac{\partial^2 \psi_{\bar{b}}(x)^2}{\partial \bar{b}^2}\bigg] \frac{\partial \bar{B}(x,x';\beta)}{\partial x} \frac{\partial \bar{B}(x,x';\beta)}{\partial x} \Big|_{x=x'} \\ &\geq& \nonumber M_2 \equiv \int_{\mathbb{R}}\ud x \lim_{\beta \rightarrow \infty} \bigg[\beta\frac{\partial^2 E(\phi_{\bar{b}}^0)}{\partial \bar{b}^2}-\frac{\partial^2 \psi_{\bar{b}}(x)^2}{\partial \bar{b}^2}\bigg] \frac{\partial \bar{B}(x,x';\beta)}{\partial x} \frac{\partial \bar{B}(x,x';\beta)}{\partial x} \Big|_{x=x'},
\end{eqnarray}\end{widetext}
where we used Fatou's lemma for the last inequality.  In this respect, notice that the evaluation of the hessian matrices is done at~$\bar{b}=\bar{B}(x,\beta)$ and that
\begin{equation}
\label{eq:98}
\beta \frac{\partial^2 E(\phi_{\bar{b}}^0)}{\partial \bar{b}^2}- \frac{\partial^2\psi_{\bar{b}}(x) }{\partial \bar{b}^2}
\end{equation}
is positive definite at each point~$x$. In order for the inequality to hold for arbitrarily large~$\beta$, when the energy hessian matrix also becomes positive definite, the following condition is necessary:
\begin{equation}
\label{eq:99}
\frac{\partial^2 E(\phi_{\bar{b}}^0)}{\partial \bar{b}^2} \frac{\partial \bar{B}(x,x';\infty)}{\partial x} \frac{\partial \bar{B}(x,x';\infty)}{\partial x} \Big|_{x=x'}=0 \quad \text{a.e}
\end{equation}
or otherwise, the argument of the last integral in~(\ref{eq:97}) becomes arbitrarily large on a set of strictly positive measure, which contradicts the fact that the integral is bounded on the entire low temperature range. The equality~(\ref{eq:99}) can be realized either by the a.e. vanishing of the derivatives of the optimizing coefficients as in the case studied in Section III, or by the vanishing of some normal modes of the energy hessian matrix. Consequently, 
\begin{eqnarray}
\label{eq:100}
&&\lim_{\beta \rightarrow \infty}e^{\beta E_{best}}\frac{\partial^2 }{\partial \bar{b}^2}\rho_{\bar{b}}^{a}(x;\beta) \frac{\partial \bar{B}(x,x';\beta)}{\partial x} \frac{\partial \bar{B}(x,x';\beta)}{\partial x} \Big|_{x=x'} \nonumber \\ &&=-\frac{\partial^2 \psi_{\bar{b}}(x)^2}{\partial \bar{b}^2} \frac{\partial \bar{B}(x,x';\infty)}{\partial x} \frac{\partial \bar{B}(x,x';\infty)}{\partial x} \Big|_{x=x'} \geq 0 \qquad \quad
\end{eqnarray}
and the last function integrates to $M_2\leq M_1$.
Now, we have 
\begin{eqnarray}
\label{eq:101}
&&\lim_{\beta \rightarrow \infty}\left[e^{\beta E_{best}}\frac{\partial^2}{\partial x \partial x'} \rho_{\bar{b}}^a (x,x';\beta )\right]_{x=x'} \nonumber \\ &&= \frac{\partial^2}{\partial x \partial x'} \big[\psi_{\bar{b}}(x)\psi_{\bar{b}}(x')\big]_{x=x'}
\end{eqnarray}
pointwise, but also
\begin{eqnarray}
\label{eq:102}
&&\lim_{\beta \rightarrow \infty}\int_{\mathbb{R}} \ud x\left[e^{\beta E_{best}}\frac{\partial^2}{\partial x \partial x'} \rho_{\bar{b}}^a (x,x';\beta )\right]_{x=x'} \nonumber \\ &&= \int_{\mathbb{R}} \ud x \frac{\partial^2}{\partial x \partial x'} \big[\psi_{\bar{b}}(x)\psi_{\bar{b}}(x')\big]_{x=x'}.
\end{eqnarray}
A comparison with~(\ref{eq:86}) produces
\begin{eqnarray}
\label{eq:103}
&&\lim_{\beta \rightarrow \infty}\int_{\mathbb{R}} \ud x\left[e^{\beta E_{best}}\frac{\partial^2}{\partial x \partial x'} \rho_{best}(x,x';\beta )\right]_{x=x'} \nonumber \\ &&= \int_{\mathbb{R}} \ud x \frac{\partial^2}{\partial x \partial x'} \rho_{best}^\circ(x,x')\Big|_{x=x'}+M_1-M_2.
\end{eqnarray}
With these results at hand, it is not hard to conclude that
\begin{eqnarray}
\label{eq:104}
 \lim_{\beta \rightarrow \infty}\left\langle E \right\rangle^{H}_{\beta}&=&\lim_{\beta \rightarrow \infty}\left\langle E \right\rangle^{H,1}_{\beta}+ \frac{\hbar^2}{2m}\frac{M_1}{\int_{\mathbb{R}} \rho_{best}^\circ(x)\ud x} \nonumber \\&=& E_1(\rho_{best}^\circ)+\frac{\hbar^2}{2m}\frac{M_1 -M_2}{\int_{\mathbb{R}} \rho_{best}^\circ(x)\ud x}.
\end{eqnarray} 
In an analog manner but using~(\ref{eq:87}), one proves 
\begin{eqnarray}
\label{eq:105}
 \lim_{\beta \rightarrow \infty}\left\langle E \right\rangle^{H}_{\beta}&=&\lim_{\beta \rightarrow \infty}\left\langle E \right\rangle^{H,2}_{\beta}- \frac{\hbar^2}{2m}\frac{M_1}{\int_{\mathbb{R}} \rho_{best}^\circ(x)\ud x} \nonumber \\&=& E_2(\rho_{best}^\circ)-\frac{\hbar^2}{2m}\frac{M_1 -M_2}{\int_{\mathbb{R}} \rho_{best}^\circ(x)\ud x}.
\end{eqnarray} 
Summation of~(\ref{eq:104}) and (\ref{eq:105})  produces the theorem:
\begin{6}[Semi-sum]
The low temperature limit of the H-method energy estimator is:
\begin{equation}
\label{eq:106}
 \lim_{\beta \rightarrow \infty}\left\langle E \right\rangle^{H}_{\beta}= \frac{1}{2}\left[E_1(\rho_{best}^\circ)+E_2(\rho_{best}^\circ)\right].
\end{equation}
\end{6}

Because $M_1 \geq M_2$, the various estimators introduced in this section can be put in the following order:
\begin{eqnarray}
\label{eq:107}
&& \lim_{\beta \rightarrow \infty}\left\langle E \right\rangle^{H,1}_{\beta}\leq  E_1(\rho_{best}^\circ) \leq \lim_{\beta \rightarrow \infty}\left\langle E \right\rangle^{H}_{\beta} \nonumber \\ &&\leq E_2(\rho_{best}^\circ) \leq \lim_{\beta \rightarrow \infty}\left\langle E \right\rangle^{H,2}_{\beta} \leq \lim_{\beta \rightarrow \infty}\left\langle E \right\rangle^{T}_{\beta}
\end{eqnarray}
[for the last inequality use~(\ref{eq:82}) and~(\ref{eq:95})]. For continuity reasons,  it is convenient to \emph{define} the energy of the LVP groundstate density matrix as
\begin{equation}
\label{eq:108}
E(\rho_{best}^\circ) = \frac{1}{2}\left[E_1(\rho_{best}^\circ)+E_2(\rho_{best}^\circ)\right].
\end{equation}
If the decomposition in sets~$D_i$ is true, the almost everywhere vanishing of the derivatives of the optimizing coefficients implies~$M_2=0$ and we recover~(\ref{eq:73}) as it should. But in this very important case, maybe more significant is the fact that $ E_1(\rho_{best}^\circ)$ and $ E(\rho_{best}^\circ)$ are above the true groundstate energy and therefore LVP is able to provide a variational energy which is better than $E_{best}$, the best energy predicted by the variational centroid based techniques. 

\section{The free particle reference case as the basic prototype}
To summarize the results obtained in this paper in one sentence, the dependence of the  density matrix with the coordinates $x$ and $x'$ is reproduced by the LVP approximation in a significantly better way than the dependence with the inverse temperature $\beta$. This is why the H-method estimator behaves in a better way than the T-method estimator. It is then interesting to compare the variational centroid method with the LVP method for the simple case when the reference system is the free particle one, so that $V'(x)\equiv0$. The point is that in this case we can leave any parameter optimization issues aside. 

Letting $\sigma=\sqrt{{\hbar^2\beta}/{m}} $, the LVP density matrix approximation takes the form
\begin{eqnarray}
\label{eq:119}\nonumber
\frac{\rho_{0}^{PA}(x,x';\beta)}{\rho_{fp}(x,x';\beta)}&=&\exp\left\{-\beta\int_{0}^{1} \mathbb{E}\,  V\Big[x(t)+ \sigma B_t^0 \Big]\ud t\right\}\\&=&\exp\left\{-\beta\int_{0}^{1} \overline{V}_{t,0}[x(t)]\ud t\right\},
\end{eqnarray}
where 
\begin{equation*}
\overline{V}_{t,0}(y)=\int_{\mathbb{R}}\frac{1}{\sqrt{2\pi\Gamma_{0}^2(t)}} \exp\left[-\frac{z^2}{2\Gamma_{0}^2(t)}\right]V(y+z) \ud z,
\end{equation*}
with~$\Gamma_0^2(t)$ defined by
\[
\Gamma_{0}^2(t)=\sigma^2 \mathbb{E}\, (B_t^0)^2 =\sigma^2 t(1-t).
\]
This is the zero order approximation of the so-called partial averaging method \cite{Dol85}.

The variational centroid expression for the diagonal centroid density matrix is \cite{Fey98}
\begin{equation}
\label{eq:120}
{\rho^c(\bar{x};\beta)}=(2\pi \sigma^2)^{-1/2}e^{-\beta K(\bar{x})},
\end{equation}
where 
\begin{equation*}
K(y)=\int_{\mathbb{R}}\frac{1}{\sqrt{2\pi\sigma^2/12}} \exp\left[-\frac{z^2}{2\sigma^2/12}\right]V(y+z) \ud z.
\end{equation*}

The question we want to answer is which one of the formulae (\ref{eq:119}) and (\ref{eq:120}) provides a better description of the physical system. To this end, notice that the spread in the partial averaging formula is on average twice as large as the one for the centroid approximation: 
\[
\int_0^1 \Gamma_0^2(t) \ud t = \sigma^2/6=2 (\sigma^2/12).
\]
This is so because the centroid position  is defined as a path average, being the unique value $\bar{x}$ around which the fluctuation 
$\int_0^1 (B_t^0-\bar{x})^2 \ud t$ 
of a path is minimized. It is therefore expected that the centroid approximation behaves in a  better way than the zero order partial averaging formula as far as the ``direct'' finite temperature partition function and the related T-method estimator are concerned. This should be true even if the analysis performed in this paper showed that the high temperature and the low temperature limits are the same for the two methods. However, this \emph{does not mean} that the centroid formula gives the better description of the system. To the contrary, we assert that by means of the H-method estimator, the zero order partial averaging formula provides the better description of the system as far as the average energy (and by integration against temperature, the ratio of the partition functions at different temperatures) is concerned. We shall present numerical evidence supporting our claims by analyzing a simple case of a periodic monodimensional potential. We choose a periodic potential because in this case the low temperature limits of both the partial averaging and the centroid formulae are well defined. Nevertheless, the reader should be aware of the fact that the free particle reference is the worst scenario for LVP as to its advantage over the equivalent centroid approximation. For the HO-LVP theory, the value of  $\Gamma_\omega^2(t)$ is controlled by the spread of the best fitting Gaussian and to a less extent by the temperature. Eventually, for low enough temperature, $\int_0^1\Gamma_\omega^2(t)\ud t$ equals the  spread of the best fitting Gaussian, but this also happens for the centroid based approximations. Therefore, the latter's advantage is diminished. 
  
 Let us consider a monodimensional periodic potential of period~$2L$ and let
\begin{equation}
\label{eq:121}
V(x)=\sum_{k \in \mathbb{Z}} v_k e^{ik\pi x/L}
\end{equation}
be its Fourier series. By the reality of the potential~$V(x)$, we have $v_{-k}=v_{k}^*$.
By Theorem~3, the  low temperature asymptotic of the zero order partial averaging density matrix is
\begin{equation}
\label{eq:122}
\rho_{0}^{PA}(x,x';\beta)\approx \exp\{-[S(x)+S(x')]\}\exp(-\beta v_0) \big/ \sqrt{2L},
\end{equation}
where 
\[
v_0=\frac{1}{2L}{\int_{-L}^{L}V(x)\ud x}
\]
is the cell average of the potential 
and where
\begin{equation}
\label{eq:123}
S(x)=\frac{2mL^2}{\hbar^2\pi^2}\sum_{k \in \mathbb{Z}, k \neq 0}\frac{v_k}{k^2}e^{ik\pi x/L}.
\end{equation}
Then, the low-temperature limit of the T-method estimator for both the centroid and the partial averaging approximations  is 
\begin{equation}
\label{eq:124}
\lim_{\beta \rightarrow \infty}\left\langle E \right\rangle^{T}_{\beta}= v_0, 
\end{equation} 
while the low temperature limit of the H-method estimator for the partial averaging formula is
\begin{equation}
\label{eq:125}
\lim_{\beta \rightarrow \infty}\left\langle E \right\rangle^{H}_{\beta}= v_0- \frac{\hbar^2}{2m} \frac{\int_{-L}^{L}\exp[-2S(x)]\;\|\nabla S(x)\|^2 \ud x}{\int_{-L}^{L}\exp[-2S(x)]\ud x}.
\end{equation} 

For a periodic potential, a state is called \emph{bounding} if its expected energy is strictly smaller than the potential average. If follows that the centroid method \emph{fails} to predict a bounding groundstate for the periodic potential. To the contrary, the zero order partial averaging formula \emph{does} predict a bounding groundstate by means of the H-estimator.
	\begin{figure}[tbp]  \centering
   \includegraphics[angle=270,width=8.5cm,clip=t]{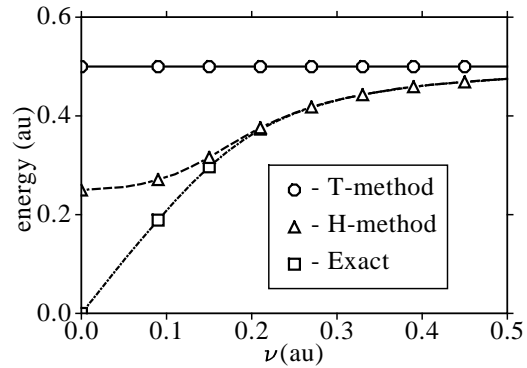} 
 \caption[sqr]
{\label{Fig:4}
The groundstate energies given by the low temperature limits of the T-method estimator and the H-method estimator are plotted together with the exact groundstate energies for various values of the potential frequency $\nu$.  
}
\end{figure}
 The predicted groundstate energy is shown in Fig.~4 for the case of the periodic potential \[V(x)=0.5\,[1+\cos(2\pi \nu x)]\] and is plotted as a function of the frequency $\nu$. Again we use atomic units and consider a particle of mass $m=1$. The exact groundstate energies were computed with the help of the Rayleigh-Ritz variational principle by expansion in a Fourier series. One notices that the H-method energy is in good agreement with the exact result in the range of high frequencies, but extrapolates to $v_0/2$ instead of $0$ in the low frequency range. 
 
 	\begin{figure}[!tbp]  \centering
   \includegraphics[angle=270,width=8.5cm,clip=t]{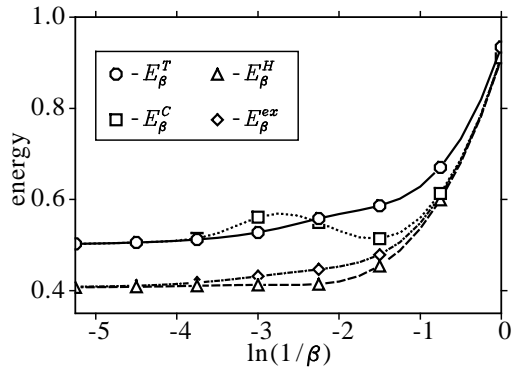} 
 \caption[sqr]
{\label{Fig:5}
Values of the partial averaging T-estimator $E^{T}_{\beta}$, partial averaging H-estimator $E^{H}_{\beta}$, centroid T-estimator $E^{C}_{\beta}$, and the exact energy $E^{ex}_{\beta}$ as functions of the inverse temperature $\beta$ in a logarithmic scale. Atomic units were used for energy and $\beta$.  
}
\end{figure} 

It is instructive to study the behavior of the finite temperature energy estimators for the periodic potential (we set $\nu =0.25$). Because in practice one stops the calculations the first time the energy fails to decrease with the temperature, Fig.~5 suggests that the centroid T-method estimator has a better behavior than the corresponding LVP T-method energy, even if their low temperature limits are the same. However, the LVP H-method estimator has an overall even better behavior, providing the closest answer to the exact energy for the  investigated range of temperatures. The exact energies were computed by numerical matrix multiplication (see for instance, Ref.~ \onlinecite{Thi83}).

\section{Concluding remarks}

	The   Local Variational Principle is alleviating some of the disadvantages of the GBF principle. Specifically, it makes better use of simple reference potentials to produce accurate diagonal density matrices. Moreover, the Trotter composition method systematically improves the finite temperature LVP density matrix up to the correct value.  Then, LVP always gives a better groundstate energy than the one predicted by the Gibbs-Bogoliubov-Feynman principle provided that the H-method estimator is used. Finally, we conclude that LVP gives a description of the physical system which is more accurate and more complete than the one provided by the centroid based approximations. I remind the reader that this situation was caused by the fact that there is no practical way of defining an H-estimator for the latter methods. Since the GBF principle becomes essentially a local approximation in the centroid space, it may be possible that by modifying the centroid idea, one might be able to construct a new technique inheriting the best features of the two methods. 
	
	Unfortunately, there are several drawbacks of the LVP method (all of them being shared by the variational centroid techniques).  The first problem is the need for expensive pointwise local maximization procedures. However, they can be  approximately performed as long as the estimators of the evaluated properties do not explicitly involve the derivatives of the optimizing parameters (both the T-method and the H-method estimators are ``stable'' in this respect). To understand the second undesirable feature, let us imagine that we slightly unbalance the double well potential. Then, the best Gaussian distribution that realizes the minimum of~(\ref{eq:42}) is unique and will be localized in the well of lower energy. Thus, there is a discontinuity of the method with respect to the balancing of the main wells of the potential.  LVP partially accounts for this problem with the help of a correction factor (see Eq.~\ref{eq:38a}), but the improvement is not always satisfactory. 
	
	Finally, let us notice that given a point $(x,x')$ in the configuration space, the HO-LVP technique gives explicit information about how the density matrix looks around this point on a range established by the temperature and by the quantum properties of the system. This information can be used to create optimal filters in conjunction with standard semiclassical approximations for the thermalized quantum dynamical correlation functions. In this respect, as discussed by Miller \cite{Mil01},  the semiclassical Van Vleck or Herman-Kluk propagators in the initial value representation  could be effectively used in quantum simulations for sufficiently large systems and for chemically relevant times if not for the associated ``sign problem.'' This is generated by the integration over the initial conditions and can be alleviated by use of filtering techniques, which are also advantageous because they require only local information about the density matrix for an optimal implementation. Though a final decision as to the feasibility of this idea awaits future detailed work, I anticipate that this local (analytical) information can be quantitatively furnished by the HO-LVP technique under the form of a ``built in'' filter.

\begin{acknowledgments}
I thank Professor J.D. Doll for introducing me in the subject of path integrals and for fruitful discussions on the local variational principle. Support from the National Science Foundation through awards CHE-9714970, CDA-9724347, and CHE-0095053 is gratefully acknowledged.
\end{acknowledgments}

\appendix
\section{Definition of the h-functions}

\begin{equation}
\label{eq:A1}
h_0(c)=\prod_{k=1}^{\infty}\frac{1}{\sqrt{1+c^2/k^2}}=\sqrt{\frac{\pi c}{\sinh(\pi c)}}
\end{equation}
\begin{eqnarray}
\label{eq:A2}
h_1(c,t)=\sum_{k=1}^{\infty}\frac{1}{k^2+c^2}\sin(k\pi t)^2\nonumber \\=\frac{\pi}{4c}\frac{\cosh(\pi c)-\cosh[\pi c (1-2t)]}{\sinh(\pi c)}
\end{eqnarray}
\begin{eqnarray}
\label{eq:A3}
h_2(c,t)=\sum_{k=0}^{\infty}\frac{1}{(2k+1)^2+c^2}\frac{1}{2k+1}\sin[(2k+1)\pi t]\nonumber \\= \frac{\pi}{4c^2}\left\{1-2\frac{\sinh[\pi c(1-t)]}{\sinh(\pi c)} +\frac{\sinh[\frac{\pi}{2} c(1-2t)]}{\sinh(\frac{\pi}{2} c)}\right\} \quad
\end{eqnarray}
\begin{eqnarray}
\label{eq:A4}
h_3(c,t)=\sum_{k=1}^{\infty}\frac{1}{(2k)^2+c^2}\frac{1}{2k}\sin(2k\pi t)\nonumber \\
=\frac{\pi}{4c^2}\left\{(1-2t) -\frac{\sinh[\frac{\pi}{2} c(1-2t)]}{\sinh(\frac{\pi}{2} c)}\right\} \quad
\end{eqnarray}
\begin{equation}
\label{eq:A5}
h_4(c)=\sum_{k=1}^{\infty}\frac{1}{k^2+c^2}=\frac{\pi}{2c}\frac{\cosh(\pi c)}{\sinh(\pi c)}-\frac{1}{2c^2}
\end{equation}
\begin{eqnarray}
\label{eq:A6}
h_5(c)=\sum_{k=0}^{\infty}\frac{1}{[(2k+1)^2+c^2]^2}\nonumber \\=\frac{\pi}{4c^2}\left[\pi\frac{\cosh(\pi c)^2}{\sinh(\pi c)^2}-\pi +\frac{1}{c}\frac{\cosh(\pi c)}{\sinh(\pi c)} \right] \nonumber \\
-\frac{\pi}{8c^2}\left[\frac{\pi}{2}\frac{\cosh(\frac{\pi}{2} c)^2}{\sinh(\frac{\pi}{2} c)^2}-\frac{\pi}{2} +\frac{1}{c}\frac{\cosh(\frac{\pi}{2} c)}{\sinh(\frac{\pi}{2} c)} \right]
\end{eqnarray}
\begin{eqnarray}
\label{eq:A7}
h_6(c)=\sum_{k=1}^{\infty}\frac{1}{[(2k)^2+c^2]^2}\nonumber \\=
\frac{\pi}{8c^2}\left[\frac{\pi}{2}\frac{\cosh(\frac{\pi}{2} c)^2}{\sinh(\frac{\pi}{2} c)^2}-\frac{\pi}{2} +\frac{1}{c}\frac{\cosh(\frac{\pi}{2} c)}{\sinh(\frac{\pi}{2} c)}-\frac{4}{\pi c^2} \right]
\end{eqnarray}
\begin{equation}
\label{eq:A8}
h_7(c,t)=\frac{\sinh[\pi c (1-t)]}{\sinh(\pi c)}\frac{\sinh[\frac{\pi}{2} c(1-2t)]}{\sinh(\frac{\pi}{2} c)}
\end{equation}

\end{document}